\documentclass{article}

\usepackage{arxiv}

\usepackage[utf8]{inputenc} 
\usepackage[T1]{fontenc}    
\usepackage{hyperref}       
\usepackage{url}            
\usepackage{booktabs}       
\usepackage{amsfonts}       
\usepackage{nicefrac}       
\usepackage{microtype}      
\usepackage{enumitem}       
\usepackage{lipsum}         
\usepackage{graphicx}
\usepackage[numbers]{natbib}
\usepackage{doi}
\usepackage{tabularx}
\usepackage{array}
\usepackage{amsmath}
\usepackage[ruled, vlined, linesnumbered]{algorithm2e}
\usepackage{multirow} 
\usepackage[most]{tcolorbox}
\definecolor{myblue}{HTML}{1C35B3}
\usepackage[caption=false,font=footnotesize]{subfig}
\usepackage{cleveref}       

\newcolumntype{Y}{>{\centering\arraybackslash}X}
\newcolumntype{C}{>{\centering\arraybackslash}X}
\title{Self-Evolving Multi-Agent Framework for Efficient Decision Making in Real-Time Strategy Scenarios}

\date{} 

\newif\ifuniqueAffiliation
\uniqueAffiliationtrue

\ifuniqueAffiliation 
\author{%
\parbox{\textwidth}{%
\centering
\textbf{Lin MA}$^1$, \enspace
\textbf{Hao PENG}$^{1}$\thanks{Corresponding author: \texttt{penghao@buaa.edu.cn}}, \enspace
\textbf{Yiming WANG}$^1$, \enspace
\textbf{Jie LIU}$^{2,3}$, \enspace
\textbf{Kongjing GU}$^4$, \\[1mm]
\textbf{Kangli ZI}$^{5,6}$, \enspace
\textbf{Na HAN}$^7$, \enspace
\textbf{Hui LIN}$^8$ \\[3mm]
{\normalfont\small
$^1$School of Cyber Science and Technology, Beihang University, Beijing 100191, China \\[-0.2mm]
$^2$Naval Aviation University, Yantai 264001, China \\[-0.2mm]
$^3$Military Science Academy, Beijing 100091, China \\[-0.2mm]
$^4$National University of Defense Technology, Changsha 410073, China \\[-0.2mm]
$^5$State Key Laboratory of AI Safety, Institute of Computing Technology, Chinese Academy of Sciences, Beijing 100190, China \\[-0.2mm]
$^6$University of Chinese Academy of Sciences, Beijing 100049, China \\[-0.2mm]
$^7$School of National Security, People's Public Security University of China, Beijing 100038, China \\[-0.2mm]
$^8$China Academy of Electronics and Information Technology, Beijing 100846, China
}%
}%
}
\else
\usepackage{authblk}

\newbox{\orcid}\sbox{\orcid}{\includegraphics[scale=0.06]{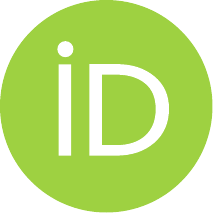}} 
\author[1]{%
	\href{https://orcid.org/0000-0000-0000-0000}{\usebox{\orcid}\hspace{1mm}David S.~Hippocampus\thanks{\texttt{hippo@cs.cranberry-lemon.edu}}}%
}
\author[1,2]{%
	\href{https://orcid.org/0000-0000-0000-0000}{\usebox{\orcid}\hspace{1mm}Elias D.~Striatum\thanks{\texttt{stariate@ee.mount-sheikh.edu}}}%
}
\affil[1]{Department of Computer Science, Cranberry-Lemon University, Pittsburgh, PA 15213}
\affil[2]{Department of Electrical Engineering, Mount-Sheikh University, Santa Narimana, Levand}
\fi

\renewcommand{\shorttitle}{}

\begin{document}
\maketitle

\begin{abstract}
    Large language models (LLMs) have demonstrated exceptional potential in complex reasoning, pioneering a new paradigm for autonomous agent decision making in dynamic settings. However, in Real-Time Strategy (RTS) scenarios, LLMs suffer from a critical speed-quality trade-off. Specifically, expansive state spaces and time limits render inference delays prohibitive, while stochastic planning errors undermine logical consistency. To address these challenges, we present SEMA (\textbf{S}elf-\textbf{E}volving \textbf{M}ulti-\textbf{A}gent), a novel framework designed for high-performance, low-latency decision-making in RTS environments. This collaborative multi-agent framework facilitates self-evolution by adaptively calibrating model bias through in-episode assessment and cross-episode analysis. We further incorporate dynamic observation pruning based on structural entropy to model game states topologically. By distilling high dimensional data into core semantic information, this approach significantly reduces inference time. We also develop a hybrid knowledge-memory mechanism that integrates micro-trajectories, macro-experience, and hierarchical domain knowledge, thereby enhancing both strategic adaptability and decision consistency. Experiments across multiple StarCraft II maps demonstrate that SEMA achieves superior win rates while reducing average decision latency by over 50\%, validating its efficiency and robustness in complex RTS scenarios.
\end{abstract}


\section{Introduction}
Developing autonomous agents equipped with profound strategic reasoning and real-time adaptive capabilities remains a cornerstone challenge in current artificial intelligence research, particularly within complex and dynamic environments~\cite{1,2}. Real-time strategy (RTS) scenarios, characterized by high-dimensional state spaces and stringent second-level interaction requirements, impose rigorous demands on both logical reasoning and reactive agility. Taking StarCraft II as a representative case, the dual requirements of strategic depth and real-time responsiveness serve as a fundamental benchmark for evaluating an agent's proficiency, ranging from macro-strategic planning to micro-tactical execution~\cite{3}. Traditional Reinforcement Learning (RL) is inherently constrained by prohibitive training costs and limited interpretability~\cite{4}. Recently, the emergence of Large Language Models (LLMs) has introduced a new paradigm for real-time decision-making~\cite{5}. Their extensive knowledge reserves and sophisticated reasoning capabilities demonstrate significant potential for addressing complex strategic maneuvers~\cite{6, 7},  yet are still beset by fundamental technical limitations and challenges in practical scenarios.

To address the formidable challenges in RTS decision-making, traditional methodologies mainly rely on rule-based algorithms~\cite{8}. However, these approaches are constrained by rigid, predefined logic, resulting in significantly limited generalization capabilities within dynamically evolving and unseen scenarios. Alternatively, RL~\cite{9} has been employed, but it typically requires a pruned action space and meticulously designed reward functions, while also facing prohibitive computational costs associated with large-scale training and convergence stability issues. In recent years, researchers have begun exploring the integration of LLMs with advanced decision-tree search to enhance path-finding capabilities within contexts involving complex multi-step logic~\cite{10}. 

As the field gradually shifts toward RTS inferencing driven exclusively by LLMs, relevant research has bifurcated into two distinct directions: Reasoning and Decision Making. The former emphasizes long-horizon planning and strategic logic, leveraging the model's intrinsic prior knowledge to formulate global strategies~\cite{11}; however, it is often prone to logical latency when capturing transient environmental changes~\cite{12}. The latter focuses on providing immediate interactive responses, requiring real-time, action-level tactical feedback based strictly on raw high-dimensional state information retrieved directly from the dynamic environment in time~\cite{13}. 

\begin{figure}[t!] 
    \centering 
    \includegraphics[width=\textwidth]{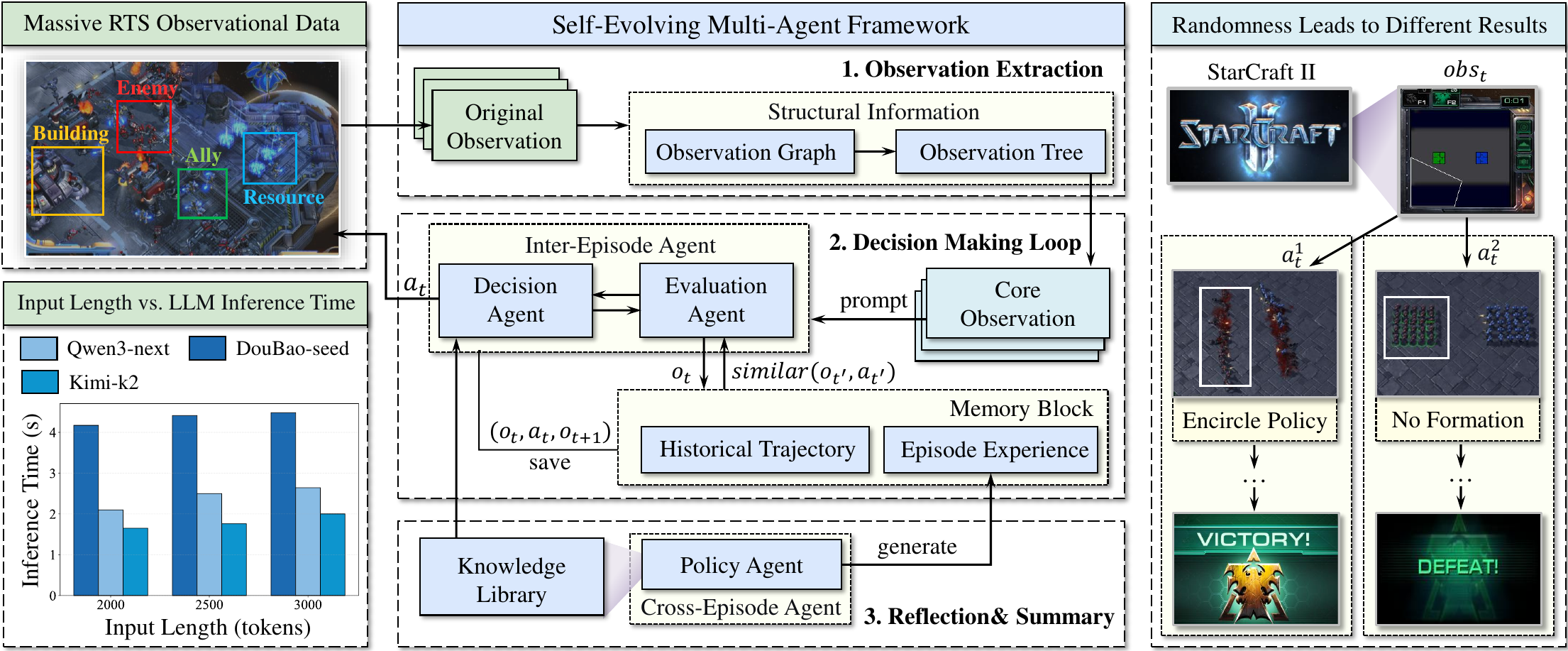} 
    \caption{The SEMA framework addresses two pivotal challenges for LLMs in RTS environments. First, the massive observational data leads to excessive input sequences, escalating reasoning latency and hindering real-time response. Second, the inherent stochastic bias of LLMs induces inconsistent decision logic, even in the exact same scenario, two distinct decisions can yield diametrically opposed outcomes, such as a shift from victory to defeat. This volatility severely undermines the robustness of agents in complex adversarial settings.}
    \label{Fig.framework} 
\end{figure}

However, direct decision-making by LLMs is severely hindered by the massive influx of high-dimensional observational data. This redundancy generates prohibitive computational latency and fails to meet the rigorous temporal demands of dynamic environments. Furthermore, the intrinsic stochasticity of LLMs can introduce undesirable and persistent logic inconsistencies, significantly undermining the overall reliability of the generated strategies. Consequently, there remains a critical void in robust frameworks capable of achieving the low-latency, high-quality responses essential for competitive play.  As illustrated in Figure \ref{Fig.framework}, raw environmental data are characterized by high dimensionality and substantial informational redundancy. This explosion of observations leads to catastrophic LLM token overflow, which has emerged as a critical technical bottleneck for real-time inference. Since the input context length and complexity directly dictate the overall inference efficiency, excessive overhead not only consumes vast computational resources but also triggers irreversible decision failure due to accumulated response lag~\cite{14, 15}. Consequently, the crux of enhancing LLM-based real-time decision-making lies in achieving effective feature refinement and adaptive structured, topological dimensionality reduction—distilling high-dimensional observations without compromising the essential semantic integrity required for complex strategic reasoning.

To surmount these limitations, we design a multi-agent collaborative framework termed SEMA based on LLMs, which features strong decision-making and reasoning capabilities, enabling rapid responses in dynamic RTS scenarios without the need for extensive fine-tuning or training. This framework constructs a self-organizing collaborative evolution system composed of a decision-making agent, an evaluation agent, and a policy agent, achieving a robust closed-loop interaction with the RTS environment. First, during the data preprocessing phase, we introduce a novel dynamic pruning mechanism for observational information based on structural entropy, which systematically structures and semantically compresses high-dimensional data, significantly reducing the computational cost of inference for LLMs. At the execution level, the policy agent updates the hierarchical experience pool to achieve cross-round strategy evolution through rigorous post-game summarization and reflection. The evaluation agent provides timely action corrections and strategic guidance based on historical trajectories through real-time retrieval of memory blocks and the experience pool, thus effectively suppressing the stochastic bias of the model. Each step in the collaborative workflow of this framework is transparent, offering clear interpretability for complex and multi-faceted decision-making processes.

We evaluated SEMA on multiple maps in StarCraft II, the results and related analysis demonstrate the superiority of the framework. Compared to HIMA~\cite{16}, our approach, based on structural information theory, substantially reduces redundant information by pruning observational data, decreasing input tokens by 70\% and reducing the response time to decisions by 50\%. Further experimental analysis indicates that this multi-agent collaborative framework broadens the tolerance space for errors, while the rapid response provides ample opportunity and time for subsequent correction and adjustment to a certain extent.

In summary, the main contributions are as follows:
\begin{itemize}[wide=\parindent, labelsep=0.5em, nosep]
    \item We propose a novel multi-agent collaborative framework termed SEMA for decision making in RTS scenarios. To our knowledge, we are the first to utilize LLMs for real-time inference without fine-tuning. 
    
    \item A structural entropy-driven pruning mechanism is formulated to eliminate semantic redundancy through structured modeling and dynamic importance updates.

    \item An enhanced learning agent is developed to achieve logical consistency between long-horizon strategies and short-term execution, leveraging historical experiences and hierarchical knowledge for self-evolution.

    \item Experimentally, SEMA achieves up to a 100\% win rate across multiple StarCraft II maps and difficulty levels. Compared to existing LLM frameworks, it reduces decision latency by over 50\%.
    
\end{itemize}

\section{Related Work}
The decision-making paradigm of multi-agent systems (MAS) is undergoing a profound evolution from static modeling based on specific tasks to a general collaborative architecture with LLMs as the cognitive core. In early classical theory, Wooldridge et al. \cite{17} defined the autonomy and sociality that agents should possess, whereas contemporary research achieves higher-level semantic synergy through LLMs. The MetaGPT framework proposed by Hong et al. \cite{18} successfully decomposes complex tasks into multi-agent circulation processes with professional roles by introducing Standard Operating Procedures (SOPs). ChatDev, developed by Qian et al. \cite{19}, utilizes the communicative agents paradigm to significantly enhance the robustness of collective intelligence through iterative interactions between agents. To explore autonomous alignment mechanisms among agents, Li et al. \cite{20} proposed the Camel role-playing framework, establishing a foundational communication protocol for autonomous collaboration between LLMs. In dynamic adversarial and gaming scenarios, MindAgent by Gong et al. \cite{21} investigates the balance between individual decision-making and global collaboration in large-scale unit confrontations. Furthermore, Furthermore, Park et al. \cite{22} utilized the Generative Agents architecture to showcase the potential of long-term memory in facilitating autonomous agent interactions and dynamic environment adaptation. Regarding communication overhead, Zeng et al. \cite{23} explored paths for semantic compression to reduce communication costs while maintaining decision consistency. To address system evaluation challenges, Zhang et al. \cite{24} proposed a framework for task decomposition and autonomous reflection in multi-agent systems. In terms of system collaborative evolution, RMIOv2 \cite{25} presents a pioneering model-based reinforcement learning framework for multi-agent systems, introducing cross-agent Transformer fusion and masked fine-tuning to maintain robust coordinated decision-making even under transient observation loss and environmental interference. These studies collectively constitute the theoretical cornerstone of current LLM-based MAS.

Planning and reasoning provide the logical framework for long-range tasks. Since Wei et al. \cite{26} introduced Chain-of-Thought (CoT) prompting, reasoning enhancement techniques have become a core means of improving the logicality of LLMs. Tree of Thoughts (ToT) by Yao et al. \cite{27} and Graph of Thoughts (GoT) by Besta et al. \cite{28} reinforce the path-planning capabilities of agents in complex decision trees by introducing heuristic search strategies. OpenVLA by Kim et al. \cite{29} and the CLEA framework by Lei et al. \cite{30} illustrate how large-scale vision-language-action paradigms can map high-level semantic abstractions onto the space of physical feasibility. Specifically, while OpenVLA provides a foundation for direct policy synthesis, CLEA further enhances long-horizon task execution through a closed-loop mechanism tailored for maintaining operational realizability within dynamic environments. For long-range dependency tasks, the DEPS framework by Wang et al. \cite{31} mitigates logic failure through hierarchical planning. Furthermore, inspired by classical game searches \cite{32, 33}, recent studies such as MCTS-Judge by Wang et al. \cite{34}, world model planning by Hao et al. \cite{35}, and the reasoning evaluation framework by Saha et al. \cite{36} all point toward a technical trend of trading test-time scaling for planning accuracy. To enhance cognitive depth in complex environments, Xi et al. \cite{37} discussed the significant potential of agent societies in handling complex tasks. Wan et al. \cite{38} proposed the Reasoning-Aware Self-Consistency (RASC) framework, which optimizes sampling efficiency and reasoning faithfulness in real-time environments by dynamically evaluating reasoning paths to navigate the efficiency-accuracy trade-off. However, multi-step reasoning strategies are often accompanied by high computational overhead, making it difficult to meet second-level decision-making requirements, which prompts researchers to turn toward more efficient LLM general decision-making architectures.

Employing LLMs for coherent decision logic is fundamental to strategic task execution. Unlike reward-based reinforcement learning \cite{39}, LLMs process complex semantic observations directly. Voyager \cite{40} enables self-evolving decision-making via skill libraries, while Agent-Omni \cite{41} enhances multimodal perception through collaborative test-time reasoning. LLMs also demonstrate generalization in autonomous driving \cite{42} and spatial understanding when integrated with geometric learning \cite{43}. To counter logic drift and hallucinations, To counter logic drift and hallucinations, STeCa \cite{44} and PT-ALIGN \cite{45} introduce step-level trajectory refinement and dual self-alignment mechanisms for autonomous reasoning calibration. From a game-theoretic perspective, research has revealed LLM behavioral biases \cite{46} and and optimized decision-making through option-regularized cooperative navigation \cite{47}. Furthermore, LLM-assisted formation control frameworks \cite{48} and benchmarking studies on long-horizon planning with verifiable constraints \cite{49} demonstrate the potential of LLMs in solving complex strategic tasks, and feedback learning in StarCraft II \cite{50} have significantly bolstered decision robustness.

\section{Preliminary}

\subsection{Structural Information Principles}
Structural entropy\cite{51} is a fundamental metric for quantifying the amount of information embedded within a complex network. It aims to quantitatively describe the dynamic characteristics of graph topological evolution, ranging from complete randomness to high degrees of regularity. Derived from the Structural Information Principles, its core logic lies in measuring the uncertainty of a graph under a specific hierarchical partitioning strategy, thereby revealing the dynamic complexity hidden behind intricate connections. Unlike classical information entropy, structural entropy considers not only the probability distribution of nodes but also deeply integrates the graph’s topological connectivity. It provides a unified mathematical framework for understanding community attributes, hierarchical organization, and information transmission efficiency within network structures.

\textbf{Graph Formulation.} Given a weighted undirected graph $G = (V, E, W)$, where $V$ is the set of vertices, $E$ is the set of edges, and $W$ is the weight function representing the intensity of interactions between nodes. For any node $v \in V$ in the graph, its degree $d_v$ is defined as the sum of the weights of all edges connected to it, reflecting the importance of the node in the local topology. To measure the scale of connectivity on a global level, the volume $vol(U)$ of any vertex subset $U \subseteq V$ is defined as the sum of the degrees of all nodes within that set
\begin{equation}
vol(U) = \sum_{v \in U} d_v.
\end{equation}

\textbf{Encoding Tree.} An Encoding Tree $T$ is the core mathematical operator for mapping the organizational structure of a graph within structural information principles. It establishes an abstract path from micro-nodes to macro-communities by hierarchically mapping vertices of the graph to the leaf nodes of the tree. The core attributes of an encoding tree are as follows:

\begin{itemize}[wide=\parindent, labelsep=0.5em, nosep]
    \item 
    Node Correspondence. Each node $\alpha \in T$ uniquely corresponds to a vertex subset $T_\alpha \subseteq V$ in the graph, ensuring a logical mapping between hierarchies while preserving the underlying structural integrity of the network.
    
    \item
    Root Node Constraint. The root node $\lambda$ corresponds to the full set of vertices, i.e., $T_\lambda = V$, representing the global perspective of the system and the total capacity of its structural information.
    
    \item 
    Parent-Child Partitioning Logic. For each node $\alpha$, the child nodes $\{\alpha \langle i \rangle\}$ satisfy the mapping $\alpha \langle i \rangle^- = \alpha$, ensuring that each child node has a unique and exclusive parent mapping. The associated vertex subsets are defined to be pairwise disjoint, thereby ensuring a rigorous and unambiguous partitioning of the graph’s topology.

    \item 
    Completeness Requiremen. The union of all child node sets must equal their parent node set, i.e., $T_\alpha = \bigcup_i T_{\alpha \langle i \rangle}$, ensuring the lossless transmission of information across every successive level of the hierarchical tree.

    \item 
    Leaf Node Attribute. The leaf nodes $v$ at the terminus of the encoding tree contain only a single vertex $T_v$ from the graph, serving as the minimal information unit that defines the highest resolution of the partition.

\end{itemize}

\textbf{One-Dimensional Structural Entropy.} The one-dimensional structural entropy of a graph $G$ is designed to characterize the global information abundance of the entire graph in the absence of hierarchical partitioning. Its physical connotation can be regarded as a measure of the orderliness of the network topology, quantitatively characterizing the extent to which the node degree distribution evolves from a state of random disorder with equal status toward a structurally centralized configuration where connections are highly concentrated. One-dimensional structural entropy is established directly upon the node degree distribution and is given by the following formula:

\begin{equation}
H^1(G) = -\sum_{v \in V} \frac{d_v}{vol(G)} \log_2 \frac{d_v}{vol(G)}.
\end{equation}

This metric reflects the degree of uniformity in network connection strength. When the graph $G$ is a highly uniform regular graph, the one-dimensional structural entropy tends toward its maximum value, suggesting that the system is in a state of high random disorder; conversely, when core nodes with extremely high degrees appear in the graph, the structural entropy decreases significantly, indicating that information is concentrated toward critical hubs. In practical research, one-dimensional structural entropy typically serves as a baseline reference for the total information volume of the system and is utilized to evaluate the contribution of nodes to the graph.

\textbf{$K$-dimensional Structural Entropy.} When a graph structure exhibits significant community aggregation or hierarchical characteristics, a one-dimensional description often overlooks local structural order; therefore, $K$-dimensional structural entropy is introduced to measure the residual uncertainty of the graph under a hierarchical partitioning strategy with a height of at most $K$. Given an encoding tree $T$, the $K$-dimensional structural entropy of graph $G$ with respect to $T$ is defined as
\begin{equation}
H^K(G) = \min_T \left\{ \sum_{\alpha \in T, \alpha \neq \lambda} H^T(G; \alpha) \right\}.
\end{equation}
The entropy term $H^T(G; \alpha)$ contributed by each non-root node $\alpha$ in the encoding tree $T$ is calculated as
\begin{equation}
H^T(G; \alpha) = -\frac{g_\alpha}{vol(G)} \log_2 \frac{V_\alpha}{V_{\alpha^-}}.
\end{equation}

In the above expression, $g_\alpha$ represents the cut weight of the subtree $T_\alpha$ rooted at $\alpha$ interacting with the external environment, which is used to quantify the cross-boundary information flow between layers; $V_\alpha$ and $V_{\alpha^-}$ denote the volumes of the subtree node set and its parent node set, respectively, measuring the relative proportion of local organization within the global perspective. By minimizing the combinatorial search over encoding trees, $K$-dimensional structural entropy can effectively map out the multi-scale hierarchical correlations of the graph, thereby achieving precise modeling and feature extraction of topological information across multiple scales.

\subsection{Formal Expression}
Referencing the Partially Observable Markov Decision Process (POMDP)\cite{52}, we formalize the RTS decision-making task as a five-tuple $\langle S, A, T, \Omega, R \rangle$, where $S$ represents the global state space comprising resource reserves, unit coordinates, and terrain layouts, and $A$ denotes the action space encompassing operations such as movement, attack, and construction. The evolution of the environment is described by the state transition function $P(s_{t+1}|s_t, a_t)$, while due to the fog-of-war characteristics of RTS environments, the agent at time $t$ can only obtain a limited local observation $o_t \in \Omega$ through the observation function $O(s_t)$. To guide policy optimization, the reward function $R(s, a)$ defines instantaneous returns based on metrics such as win-loss outcomes and kill-loss ratios.

\begin{table}[t!]
\centering
\footnotesize
\caption{List of Notations for SEMA Framework.}
\label{tab:notations}
\begin{tabularx}{\textwidth}{lX|lX|lX} 
\toprule
\textbf{Notation} & \textbf{Definition} & \textbf{Notation} & \textbf{Definition} & \textbf{Notation} & \textbf{Definition} \\ \midrule
$G$      & Graph structure            & $o$      & Local observation         & $\Phi$   & Spatio-temporal operator \\
$V$      & Vertex set                 & $\mathcal{N}$ & Multi-agent set       & $w$      & Normalized edge weight   \\
$E$      & Edge set                   & $\mathcal{L}$ & Hierarchical storage  & $\Delta$ & Attribute variation      \\
$W$      & Global weight function     & $\mathcal{M}$ & Micro-memory block    & $\delta$ & Variational metric       \\
$d$      & Node degree                & $\mathcal{E}$ & Macro-experience pool & $\mu$    & Pruning threshold        \\
$T$      & Encoding tree              & $\mathcal{K}$ & External knowledge base & $N$      & Capacity factor          \\
$\lambda$& Root node of $T$           & $a$      & Decision command          & $\tau_{max}$ & Latency Limit        \\
$\alpha$ & Tree node                  & $\eta$   & Deviation tolerance       & $\gamma$ & Discount factor          \\
$H$      & Structural entropy         & $\pi$    & Optimal policy            & $\zeta$  & Win-loss threshold       \\
$g$      & Cut weight                 & $S$      & Global state space        & $P$      & Transition function      \\ 
$A$      & Action space               & $\Omega$ & Observation space         & $R$      & Reward function          \\ 
$\text{e}$      & Unit                       & $\text{op}$     & Action id                & $\text{ta}$      & Target                   \\ \bottomrule
\end{tabularx}
\end{table}

Building upon this foundational environment, the SEMA framework implements an extension of complex real-time decision logic by introducing specialized operators. Table \ref{tab:notations} summarizes major notations used in this paper. First, the system constructs a multi-agent set $\mathcal{N}$ synergistically composed of a decision agent $\mathcal{N_D}$, an evaluation expert $\mathcal{N_V}$, and an analysis expert $\mathcal{N_A}$, and establishes a hierarchical storage structure $\mathcal{L} = \langle \mathcal{M}, \mathcal{E}, \mathcal{K} \rangle$. Within this structure, the micro-memory block $\mathcal{M}$ stores step-level trajectory tuples to support tactical reuse during high-frequency combat scenarios, the macro-experience pool $\mathcal{E}$ stores strategic abstractions extracted by $N_A$, and the external knowledge base $\mathcal{K}$ provides static domain rules; together, these three components constitute the foundation for multi-scale knowledge enhancement. Simultaneously, addressing the stochastic errors prone to LLMs, we define the optimal action under ideal logic as $a^*_t$ and establish a logical consistency constraint $\| a_t - a_t^* \| \le \eta$ via the deviation tolerance $\eta$.

In summary, the core scientific problem of this research can be summarized as: how to derive the optimal decision sequence $\tau = \{a_1, a_2, \dots, a_T\}$ to maximize the expected global win rate under the mapping of cumulative returns, while satisfying the stringent real-time response latency constraint $\Delta t \le \tau_{max}$, through the effective retrieval and integration of the hierarchical knowledge base $\mathcal{L}$. SEMA aims to solve for the optimal policy $\pi^*$, with the objective function defined as:
\begin{equation}
\pi^* = \arg\max_{\pi} \mathbb{E}_{\tau \sim \pi} \left[ \mathbb{I} \left( \sum_{t=1}^{T} \gamma^{t-1} r_t > \zeta \right) \right],
\end{equation}
where $\mathbb{I}(\cdot)$ denotes the indicator function representing the binary win-loss outcome, $\gamma \in [0, 1]$ is the discount factor, and $r_t$ is the instantaneous reward defined by metrics such as kill-loss ratios and resource accumulation at time $t$. The goal is to simultaneously balance real-time responsiveness and logical consistency within complex dynamic games, thereby suppressing behavioral deviations induced by model stochasticity.

\section{Method}
\subsection{Framework Overview}
    \begin{figure}[t!] 
		\centering 
		\includegraphics[width=\textwidth]{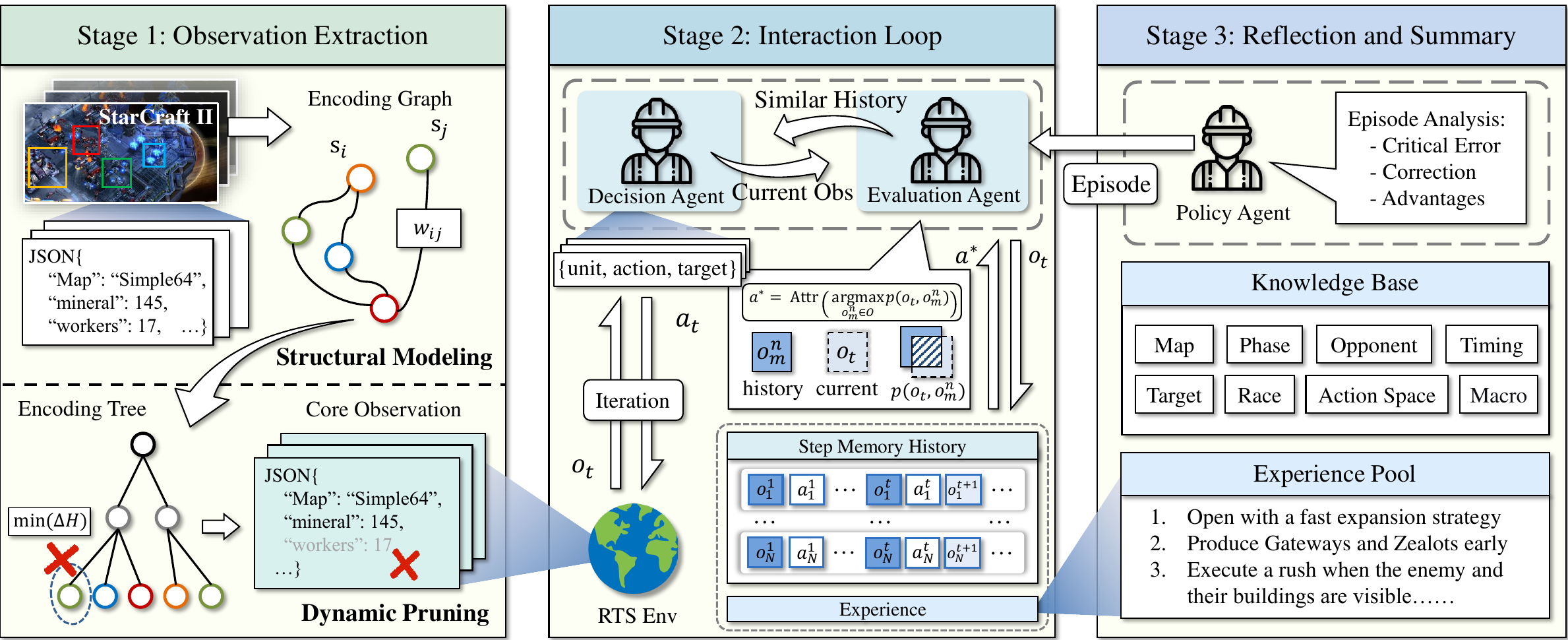} 
        \captionsetup{justification=raggedright,singlelinecheck=false}
		\caption{Overview of SEMA. First, structural modeling and dynamic pruning are employed to extract core observations, reducing reasoning latency. Second, decision and evaluation agents perform closed-loop calibration via history retrieval to suppress stochastic bias. Finally, the policy agent analyzes episode performance and updates experience, driving the continuous self-evolution of strategic logic.}
		\label{Fig./SEMA pipline} 
	\end{figure}
As shown in Figure \ref{Fig./SEMA pipline}, the SEMA framework is designed to effectively address the challenges of inference latency and logical consistency of LLMs in RTS environments by constructing a closed-loop, self-evolving multi-agent collaborative framework. It comprises three core stages: structural entropy-driven observation pruning, adaptive decision execution, and dual-loop self-evolution. Specifically, in the observation preprocessing stage, inspired by structural information theory, SEMA employs dynamic pruning techniques based on structural entropy to perform topological modeling of high-dimensional game states, significantly reducing input redundancy by extracting core semantic information. In the decision execution stage, SEMA operates through a multi-agent collaborative architecture consisting of decision, evaluation, and analysis-reflection agents, integrated with a memory-knowledge enhancement mechanism that fuses micro-trajectories and macro-experience to ensure decision consistency in complex environments. In the self-evolution stage, SEMA establishes a nested feedback loop consisting of step-level in-game assessment and episode-level post-game analysis, dynamically optimizing strategies by continuously calibrating model bias.

\subsection{Dynamic Pruning of Observation}
In RTS scenarios, the raw observation space is characterized by high dimensionality, nonlinearity, and significant redundancy. To enable efficient decision-making inference for Large Language Models, we propose a Structural Entropy-based Dynamic Observation Pruning Mechanism, which performs deep semantic compression of environmental states through a logical pipeline of topological modeling, structural information measurement, and dynamic pruning. This mechanism significantly reduces the input length for LLMs.

\textbf{Structured Modeling of Observational Information.} Given the advanced reasoning and semantic parsing capabilities of LLMs, we employ the LLM to perform the mapping and establish the edge relationships for graph construction. We map the raw observation space at time $t$ into a dynamic attribute graph $G_t = (V, E)$, where the vertex set $V$ denotes environmental attributes such as resource reserves, unit status, and construction progress, while the edge set $E$ characterizes the interactive correlations or logical dependencies between these attributes. To capture latent correlations and the evolutionary characteristics of multidimensional states, we define a spatio-temporal evolution operator $\Phi$ to measure the binary correlation intensity of attribute variations $\Delta$ between consecutive decision frames:
\begin{equation}
\Phi(\Delta_i, \Delta_j) = |\Delta_i \cdot \Delta_j| \cdot \ln(1 + |\Delta_i \cdot \Delta_j|).
\end{equation}
This operator employs a non-linear enhancement mechanism for feature scaling, enabling the detection of subtle collaborative fluctuations among attributes. This design stems from the intuition that in RTS environments, core strategic value is typically embedded in relative rates of change—such as the synergy between resource consumption and force replenishment—rather than static magnitudes. Subsequently, through global normalization, the evolution operator is projected onto a standard metric space to construct topological edge weights:
\begin{equation}
w_{ij} = \frac{\Phi(\Delta_i, \Delta_j)}{\|\mathbf{W}_t\|_{\infty}},
\end{equation}
where $\|\mathbf{W}_t\|_{\infty}$ denotes the $L_{\infty}$ norm of the current evolution matrix. This procedure yields a weighted undirected graph that facilitates the real-time representation of dynamic battlefield evolution within the feature space.

\textbf{Structural Entropy Guided Adaptive Pruning.} The pruning mechanism is designed to map the observation space onto an optimal low-dimensional representation while preserving critical semantic information. We embed the dynamic attribute graph $G_t$ into a hierarchical encoding tree $T$ of depth $K$ via a greedy evolutionary clustering algorithm. During this process, the system executes an entropy-reduction-driven node aggregation strategy, iteratively searching for the merging operator $\sigma: \{v_i, v_j\} \to X_k$ that maximizes the local structural entropy reduction. This approach aggregates attributes with strong correlations from leaf nodes into higher-order semantic communities $X \in \mathcal{P}$, thereby eliminating redundancy at the structural level. To quantitatively evaluate the contribution of each attribute node to system complexity, we construct a Structural Entropy Perturbation Variational Metric:
\begin{equation}
\delta_H = \alpha \cdot \Delta H_G + \beta \cdot \Delta H_T,
\end{equation}
where $\Delta H_G$ measures the impact of node removal on the graph’s one-dimensional structural entropy, and $\Delta H_T$ assesses the feature representativeness of a node within its local semantic cluster based on two-dimensional structural entropy $H^2(G;T)$. Through this multi-dimensional information measure, the system can precisely identify non-core attributes whose information content falls below the threshold of semantic significance.

\textbf{Dynamic Cyclic Update.} A closed-loop update cycle between the multi-agent framework and the RTS environment maintains real-time perceptual timeliness. Before each decision, the system reconstructs the graph topology from instantaneous state changes and triggers dynamic entropy evolution. To balance information compression and perceptual integrity, the pruning process is governed by the synergetic constraints of a dynamic threshold $\mu$ and a capacity factor $N$. SEMA first identifies a candidate removal set $V_{cand}$   satisfying  $\delta_{H,v} < \mu$; if the size of the candidate set exceeds the preset capacity $N$, a priority sorting algorithm is activated to selectively prune the top $N$ non-core dimensions with the smallest $\delta_H$ metrics, ensuring the robustness of the perceptual boundary. Experimental results demonstrate that while ensuring high-fidelity retention of critical strategic semantics, this mechanism achieves a significant reduction of approximately 70\% in the LLMs' token load and shortens decision response time over 50\%, markedly enhancing the inference efficiency and capabilities of LLMs in complex real-time environments.

\begin{algorithm}[H]
\IncMargin{1.5em} 
\SetNlSty{textnormal}{}{.} 
\caption{Dynamic Observation Pruning via Structural Entropy}
\label{alg:pruning}

\KwIn{Current and previous observations $O_t, O_{t-1}$; weighting factors $\alpha, \beta$; threshold $\mu$; capacity constraint $N$.}
\KwOut{Refined semantic state $O_{\text{pruned}}$.}

\BlankLine
Initialize node set $V$ from $O_t$\;

\For{each property pair $(i, j) \in V$}{
    $\Delta_i \leftarrow |O_t[i] - O_{t-1}[i]|$\;
    $\Phi_{ij} \leftarrow |\Delta_i \Delta_j| \ln(1 + |\Delta_i \Delta_j|)$\;
    $w_{ij} \leftarrow \Phi_{ij} / \|\mathbf{W}_t\|_{\infty}$\;
}

\BlankLine
Construct hierarchical encoding tree $T$ via greedy entropy minimization\;
$H^1_{\text{base}} \leftarrow H^1(G_t)$, $H^2_{\text{base}} \leftarrow H^2(G_t; T)$\;

\For{each node $v \in V$}{
    $\Delta H_G \leftarrow |H^1_{\text{base}} - H^1(G_t \setminus \{v\})|$\;
    $\Delta H_T \leftarrow |H^2_{\text{base}} - H^2(G_t \setminus \{v\}; T)|$\;
    $\delta_{H,v} \leftarrow \alpha \cdot \Delta H_G + \beta \cdot \Delta H_T$ 
}

\BlankLine
$V_{\text{cand}} \leftarrow \{v \in V \mid \delta_{H,v} < \mu\}$\;

\eIf{$|V_{\text{cand}}| > N$}{
    $V_{\text{remove}} \leftarrow$ Top-$N$ nodes from $V_{\text{cand}}$ with minimum $\delta_H$\;
}{
    $V_{\text{remove}} \leftarrow V_{\text{cand}}$\;
}
\Return $O_t \setminus V_{\text{remove}}$\;
\end{algorithm}

\subsection{Decision Agent}
The Decision Agent executes the policy mapping from the feature space to the action space. Its input consists of the compressed observation $o_t$, the historical reference actions $a^*$, and the strategic experience $\mathcal{E}$. Leveraging the reasoning mechanism of a large language model, the agent performs logical deduction under the explicit constraints of action space $A$ to output a formalized tuple $a_t = \langle \text{e}, \text{op}, \text{ta} \rangle$, ensuring determinism and interpretability of decision commands in highly complex environments. Specifically, $\text{e}$ defines the execution subject or a set of units, $\text{op}$ corresponds to an element within action space $A$, and $\text{t}$ refers to the interaction object or a set of spatial coordinates. This structured representation enables precise localization across the dimensions of execution entities, behavioral logic, and interaction targets, thereby enhancing the command generation quality and logical consistency in strategic, multi-horizon collaborative planning under dynamic and uncertain conditions.

\newtcolorbox{promptbox}[1]{
    enhanced,
    colback=white,
    colframe=myblue,   
    colbacktitle=myblue, 
    coltitle=white,           
    fonttitle=\bfseries,
    arc=3pt,
    left=8pt, right=8pt,
    top=8pt, bottom=8pt,
    titlerule=0pt,            
    title=#1,                 
}

\begin{promptbox}{Prompt: Decison Making}
    \small
\textbf{System Prompt:} You are an elite StarCraft II tactical analyst and real-time decision AI. Based on \texttt{\{tactic\}}, \texttt{\{summary\}}, and \texttt{\{obs\}}, you must orchestrate macro-operations and multi-agent micro-combat. Prioritize unit production via \texttt{\{action\_dict\}} while adhering to supply limits. Implement precise maneuvers for surviving agents. Optimize formation spacing using high-ground and choke points to ensure global synergy. Continuously evaluate enemy threat priorities to maximize damage efficiency via focus fire, and dynamically adjust engagement depth to synchronize offensive pushes and retreats, preventing formation fragmentation. \\[6pt]
\textbf{User Prompt:} Current observation: \texttt{\{observation\}}; Last observation frames: \texttt{\{history\}}; Reference decisions: \texttt{\{actions\}}. Generate optimal commands strictly following the format.\\[6pt]
\textbf{Output:} Commands must be strictly formatted as a sequence of \textbf{$(u, a, t)$ triples} ($u$=\text{e}, $a$=\text{op}, $t$=\text{ta}). If $u$ or $t$ is omitted, the system resolves assignments via spatial proximity (nearest available). Output must consist solely of raw triples (e.g., \texttt{(unit\_01, 104, enemy\_02)}, \texttt{(101)}), with no explanations or headers.
\end{promptbox}

\subsection{Evaluation Agent}
The Evaluation Agent performs multi-scale situational retrieval and semantic alignment prior to decision-making, constructing a comprehensive context for the reasoning process. At each decision step $t$, the agent receives the observation vector $o_t$ processed via structural-entropy-based dynamic pruning and retrieves a similar state $o_m^n$ from the step-level trajectory memory bank based on a semantic similarity metric. Specifically, the system calculates the cosine similarity $S(o_t, o_m^n)$ between the current observation $o_t$ and each historical state in the memory bank, and selects the sample with the highest similarity through a retrieval operator $o_m^n = \arg \max_{o \in \mathcal{M}} S(o_t, o)$. Subsequently, the corresponding historical action trajectory is extracted as a transient behavioral reference according to
\begin{equation}
a^* = \text{Attr} \left( \arg\max_{o_m^n \in \mathcal{O}} p(o_t, o_m^n) \right).
\end{equation}

Simultaneously, the agent accesses an episode-level experience pool to match current game meta-information and retrieve the global descriptive prior $\mathcal{E}$. By integrating the transient action features $a^*$ with the global strategic prior $\mathcal{E}$, the agent achieves a multi-source information fusion, providing the Decision Agent with a set of prior inputs.

\subsection{Policy Agent}
The Evaluation Agent performs post-hoc assessments of battlefield keyframes and drives the evolution of domain knowledge, enabling the continuous growth of systemic experience. Based on the metadata of these keyframes, the agent extracts strategic rules and benchmark knowledge from a hierarchical domain knowledge base. By evaluating the decision sequence $\tau$, it generates an analysis $\mathcal{E}$ that encapsulates both tactical logical advantages and deficiencies. This refined, reflective experience is then integrated into the global experience pool via an incremental update mechanism, thereby enhancing policy robustness in complex adversarial environments. This closed-loop evaluation mechanism facilitates the dynamic expansion of the SEMA perception boundary, ensuring autonomous self-correction capabilities throughout the process of policy evolution.

\begin{promptbox}{Prompt: Experience Analysis and Strategy Refinement}
    \small
    \textbf{System:} You are an elite StarCraft II tactical analyst with a profound foundation in competitive theory and data diagnostics. Your task is to perform a deep post-match review of Melee map results based on the tactical database \texttt{\{tactic\}}. Focus on diagnosing imbalances between resource efficiency (e.g., mineral floating due to operational desync), army composition (unit counters and stylistic alignment), and decision quality (alignment between build orders and map topography). During evaluation, precisely summarize critical flaws and actionable improvements while highlighting tactical highlights. Finally, output the analysis report in strict accordance with the \texttt{\{shot\}} template. \\[6pt]
    \textbf{User:} The raw observation data for the current match is \texttt{\{obs\}}. Based on the aforementioned framework and constraints, generate a highly concise tactical assessment report. \\[6pt]
    \textbf{Output:} Directly present the analysis content. Any titles or redundant section explanations are prohibited. The total length must be under 200 words, ensuring compact logic and full compliance with the template.
\end{promptbox}

\section{Experiment}
\subsection{Experimental Setting}
With Qwen3-next-80b~\cite{53} integrated as the foundational model, SEMA is evaluated across eight StarCraft II maps. This selection includes four Melee maps tested at two difficulty levels and three StarCraft Multi-Agent Challenge (SMAC) maps. While Melee maps feature expansive observation and action spaces emphasizing strategic planning, SMAC maps prioritize micromanagement requiring rapid response to precise observations. Efficacy is quantified via 50 randomized trials against the built-in AI per map, using win rate and average latency as primary metrics.

\textbf{Baselines.} We compared SEMA with diverse representative methods across the RTS landscape, from traditional heuristic and RL-based solvers to cutting-edge LLM frameworks like TextStarCraft~\cite{15} and HIMA~\cite{16}.

\textbf{StarCraft II.} StarCraft II is a highly complex and competitive RTS game. In StarCraft II, combat units can execute attack actions against enemies within their engagement range. All player actions are performed under real-time constraints and partial observability, requiring players to defeat opponents by gathering resources, constructing buildings, producing units, and commanding armies. The game concludes upon reaching specific mission objectives or eliminating enemy forces, such as destroying all units, critical structures, or main command centers. At the termination of a match, statistical feedback regarding resource efficiency and unit performance is often provided to facilitate strategy optimization for future encounters. We selected a variety of maps with diverse army scales and dimensions; the specific list and descriptions of these maps are as follows:

\begin{itemize}[wide=\parindent, labelsep=0.5em, nosep]
    \item \textbf{3m.} As a micromanagement map in the SMAC benchmark, it features a small-scale skirmish between 3 allied Marines and 3 enemy Marines, designed to evaluate the agent's basic positioning and focus-fire capabilities.
    
    \item \textbf{8m.} This map scales the engagement to an 8-on-8 Marine encounter. The increased unit density requires the agent to maintain precise local micromanagement while exhibiting superior formation control.
    
    \item \textbf{25m.} A high-density infantry combat map featuring a massive 25-versus-25 lineup. The sheer scale of the units forces the agent to resolve complex targeting logic within highly congested environments.

    \item \textbf{Flat32.} A miniature 32-by-32 flat melee map characterized by an extremely short rush distance. The engagement reaches a fever pitch almost immediately after the game starts.

    \item \textbf{Flat48.} A medium-sized 48-by-48 flat melee map that eliminates topographical interference, focusing on the agent’s ability to balance economic expansion with early-stage unit production.

    \item \textbf{Flat64.} A large, open 64-by-64 melee map with significant distance between opponents. It emphasizes macro-management, long-range reconnaissance, and global strategic positioning.
    
    \item \textbf{Simple64.} A classic 64-by-64 competitive environment featuring terrain elements such as ramps and choke points. It requires the agent to possess fundamental spatial reasoning and sophisticated offensive-defensive strategies.
\end{itemize}

\textbf{Evaluation metric.} To measure the success of our framework, win rate is utilized as the core metric and a direct correlate of decision quality. Furthermore, the evaluation metrics include game duration and average response time per move to quantify the model's efficiency in real-time decision-making environments.

\begin{table*}[t!]
    \centering
    \footnotesize
    \captionsetup{singlelinecheck=off, justification=raggedright}
    \caption{Win rate(\%) and average episode time for successful games across different methods on SMAC and Melee Maps. The time duration listed represents the execution time of a single successful episode; cases where no victory was achieved are denoted by a dash.}
    \label{tab2}
    \begin{tabularx}{\textwidth}
    {
        >{\hsize=1.8\hsize}Y 
        >{\hsize=0.7\hsize}Y 
        >{\hsize=0.7\hsize}Y 
        >{\hsize=0.7\hsize}Y 
        *{8}{>{\hsize=0.7\hsize}Y} 
    }
    \toprule
    \multirow{2}{*}{Method} & \multirow{2}{*}{3m} & \multirow{2}{*}{8m} & \multirow{2}{*}{25m} & \multicolumn{2}{c}{Flat32} & \multicolumn{2}{c}{Flat48} & \multicolumn{2}{c}{Flat64} & \multicolumn{2}{c}{Simple64} \\ 
    \cmidrule(lr{-0.5em}){5-12}
    & & & & Lv.1 & Lv.2 & Lv.1 & Lv.2 & Lv.1 & Lv.2 & Lv.1 & Lv.2 \\
    \midrule
    Random        & 2/16s & 2/9s & 0/- & 0/- & 0/- & 0/- & 0/- & 0/- & 0/- & 0/- & 0/-\\
    Rule-Based    & 50/8s & 46/9s & 55/14s & 58/6m & 53/7m & 63/8m & 58/9m & 60/10m & 56/11m & 61/10m & 59/10m\\
    Single-LLM    & 40/16s & 32/12s & 34/20s & 75/5m & 72/7m & 74/6m & 73/6m & 71/7m & 70/6m & 71/7m & 69/9m\\
    TextStarCraft & 56/14s & 44/14s & 38/19s & 83/6m & 83/6m & 87/6m & 80/7m & 86/7m & 82/6m & 85/7m & 81/8m\\
    HIMA          & 66/20s & 54/19s & 10/23s & 94/6m & 89/6m & 100/7m & 94/7m & 95/6m & 89/7m & 90/7m & 85/8m\\
    \textbf{SEMA(Ours)}    & \textbf{88/10s} & \textbf{70/17s} & \textbf{68/17s} & \textbf{100/5m} & \textbf{94/5m} & \textbf{100/5m} & \textbf{95/5m} & \textbf{100/5m} & \textbf{94/5m} & \textbf{100/5m} & \textbf{100/6m}\\
    \bottomrule
    \end{tabularx}
\end{table*}

\begin{figure*}[t!]
    \centering
    \subfloat[3m]{\includegraphics[width=0.19\textwidth]{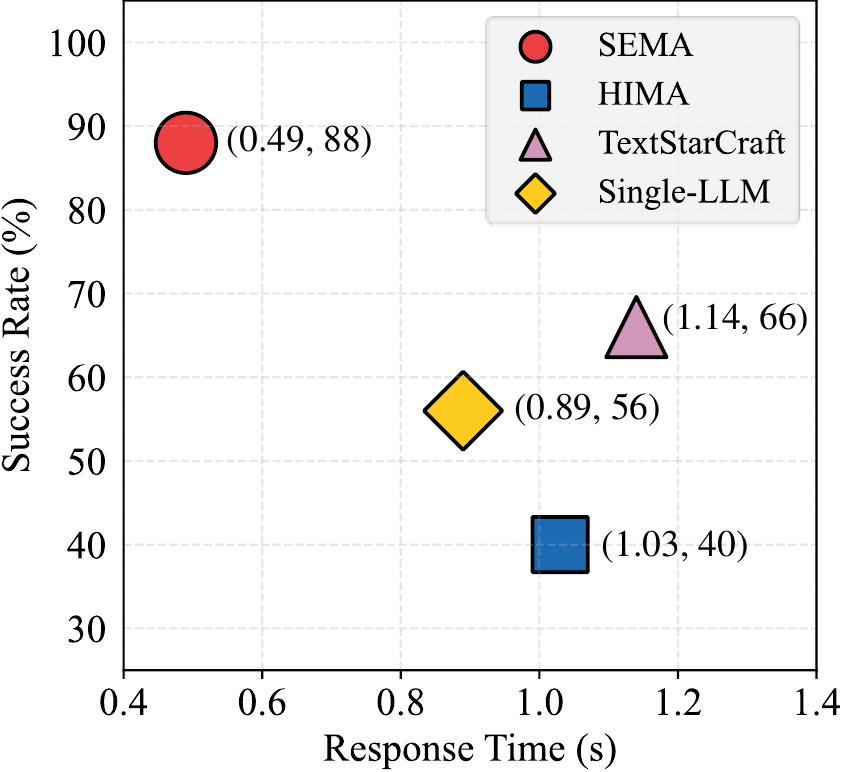}\label{fig3:a}}\hfill
    \subfloat[8m]{\includegraphics[width=0.19\textwidth]{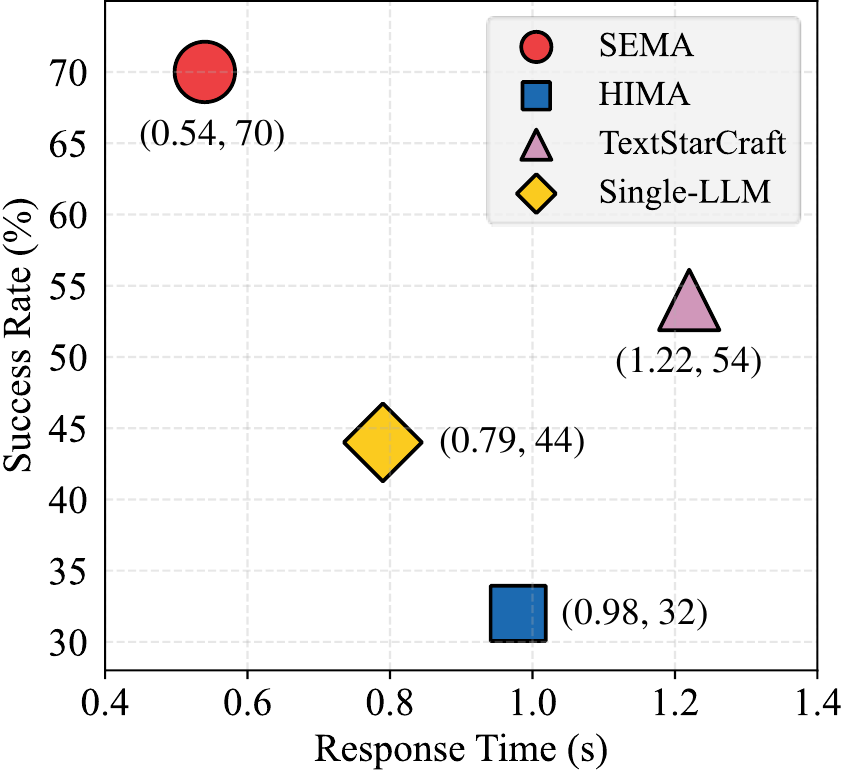}\label{fig3:b}}\hfill
    \subfloat[25m]{\includegraphics[width=0.192\textwidth]{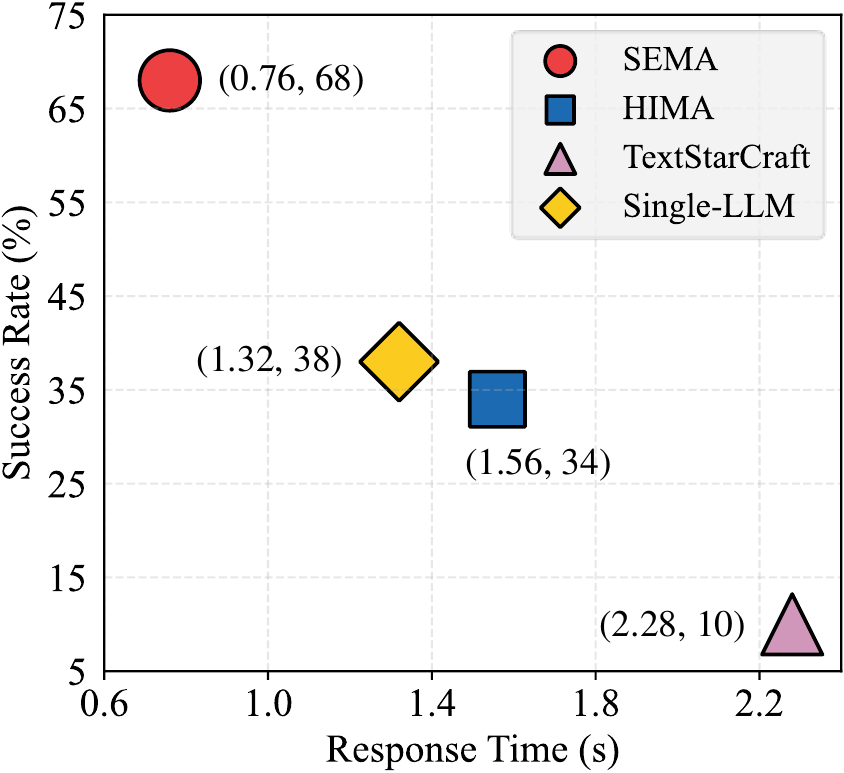}\label{fig3:c}}\hfill
    \subfloat[Flat64-Lv.1]{\includegraphics[width=0.19\textwidth]{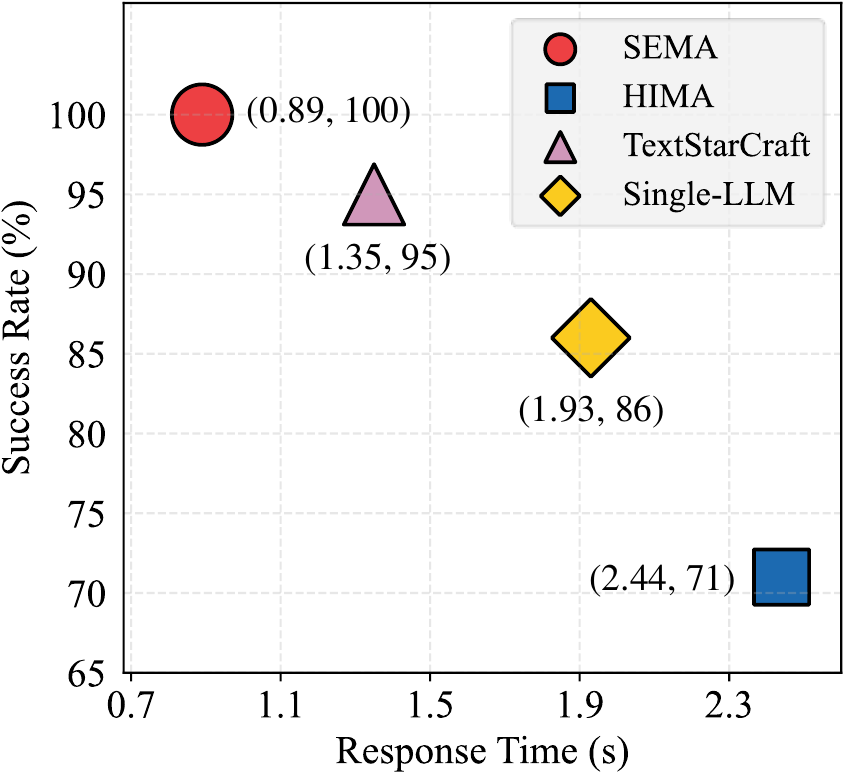}\label{fig3:d}}\hfill
    \subfloat[Flat64-Lv.2]{\includegraphics[width=0.19\textwidth]{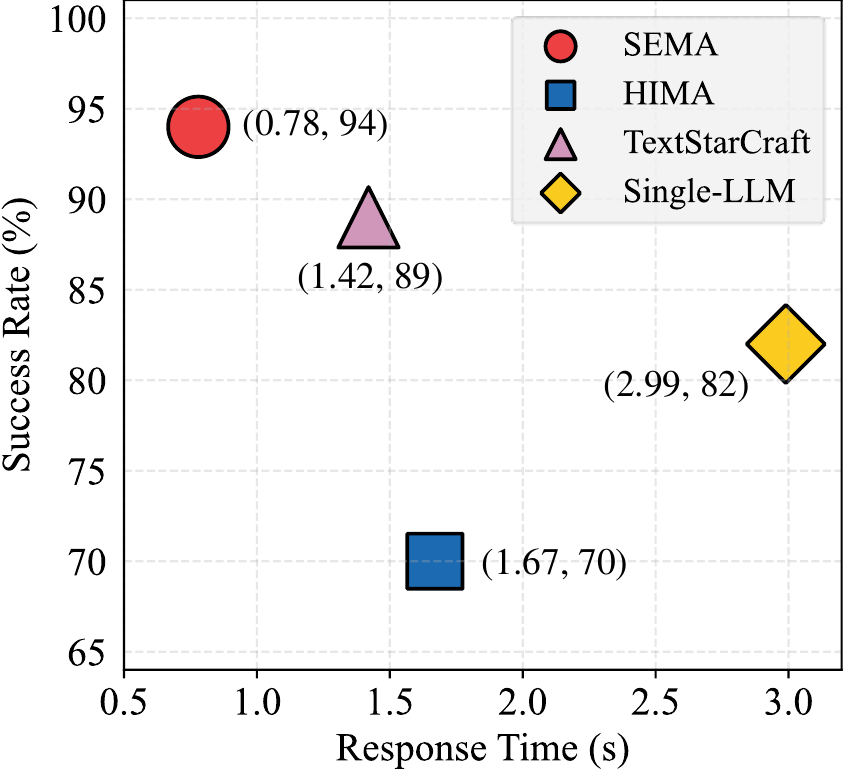}\label{fig3:e}}
    \captionsetup{singlelinecheck=off, justification=raggedright}
    \caption{Performance comparison of win rate and response time across diverse scenarios. (a) 3m; (b) 8m; (c) 25m; (d) Flat64 (VeryEasy); (e) Flat64 (Easy). For each point, the values in parentheses $(x, y)$, e.g., $(0.49, 88)$, denote the average response time $(s)$ and the success rate $(\%)$.}
    \label{Fig3}
\end{figure*}

\subsection{Experimental Result}
We first conducted extensive comparative experiments between the SEMA framework and the built-in AI system of StarCraft II across various complex scenarios. To comprehensively verify the robustness and adaptability of the framework across different tactical dimensions, the experimental design encompassed adversarial tasks on SMAC maps requiring extreme real-time micro-management, as well as tasks on Melee maps focusing on global resource management and long-term strategic planning. In all match-up configurations, our faction was consistently set as Protoss, while the opponent faction was set as Terran. On the Melee maps, we evaluated performance against the built-in AI at Lv.1 Very Easy and Lv.2 Easy difficulty levels. As shown in Table \ref{tab2}, SEMA demonstrated outstanding performance in both micro and macro test scenarios, achieving a 100\% win rate on Melee maps. These results fully demonstrate that SEMA, as a real-time decision-making framework, possesses robust strategy evolution capabilities and superior real-time reasoning proficiency.

Compared with state-of-the-art open-source LLM baselines such as HIMA and TextStarCraft, SEMA exhibits a significant dual advantage in win rate and decision efficiency. From the perspective of win rate, SEMA achieved a 68\% win rate in the highly unit-dense 25m map, which is substantially higher than the 10\% achieved by HIMA. This performance gap stems from the fact that the vast majority of the removed observational information consists of redundant data. By significantly reducing computational overhead while ensuring the integrity of core battlefield semantics, the framework enables the LLM to perform precise strategic reasoning based on state representations. Furthermore, the introduction of the self-evolving mechanism allows the system to continuously optimize itself against complex adversarial environments by iterating on historical battle experiences, thereby maintaining a prominent competitive advantage in long-horizon adversarial competitions.

The most critical breakthrough lies in the exceptional inference efficiency of SEMA, which directly dictates the decision quality and execution effectiveness in real-time confrontations. As illustrated in Figure \ref{Fig3}, the average response time of SEMA remains stable between 0.5s and 1.0s, whereas other LLM-based methods typically exhibit response times about 1.5s or even 2.0s. In rigorous tasks requiring millisecond-level responsiveness, this latency discrepancy leads to diametrically opposed outcomes. Due to excessive response delays in baseline methods, the battlefield situation often undergoes drastic changes by the time decision commands are issued. Consequently, commands are frequently applied to units that have already been eliminated or have shifted positions, resulting in significant resource waste and logical interference. In contrast, due to its rapid processing speed, SEMA ensures that tactics are executed within the current effective observation window while providing ample temporal margin for subsequent decision corrections. The experimental data corroborates our core thesis: in real-time strategy scenarios, extreme inference efficiency serves as the fundamental cornerstone for the realization of strategic reasoning.


\subsection{Ablation Study}
\begin{table}[t!]
    \centering
    \footnotesize    
    \captionsetup{singlelinecheck=off, justification=raggedright}
    \caption{Performance comparison of win rate and inference time per step across different components. The win rate serves as a key performance indicator, while the recorded time denotes the duration of a single successful episode.}
    \label{tab3}
    \begin{tabularx}{\textwidth}
    {
        >{\hsize=1.45\hsize}C 
        *{7}{>{\hsize=0.68\hsize}C} 
    }
    \toprule
    \multirow{2}{*}{Method} & \multirow{2}{*}{3m} & \multirow{2}{*}{8m}& \multicolumn{2}{c}{Flat64} & \multicolumn{2}{c}{Simple64}\\ 
    \cmidrule(lr){4-7}
    & & & Lv.1 & Lv.2 & Lv.1 & Lv.2\\
    \midrule
    w/o Dynamic Pruning    & 50 / 1.4s  & 60 / 1.4s & 90 / 1.5s & 85 / 1.9s  & 90 / 1.6s & 75 / 2.1s\\
    w/o Evaluation Agent    & 75 / 0.5s  & 67 / 0.7s & 93 / 0.7s & 86 / 0.7s & 93 / 0.6s & 88 / 0.7s\\
    w/o Policy Agent    & 47 / 0.5s  & 52 / 0.6s & 83 / 0.8s & 78 / 0.7s & 82 / 0.8s & 72  / 0.7s \\
    \textbf{SEMA (Ours)} & \textbf{88 / 0.5s} & \textbf{70 / 0.5s} & \textbf{100 / 0.9s} & \textbf{94 / 0.8s} & \textbf{100 / 0.8s} & \textbf{100 / 0.8s} \\
    \bottomrule
    \end{tabularx}
\end{table}

\begin{figure*}[t!]
\subfloat[3m]{\includegraphics[width=0.23\textwidth]{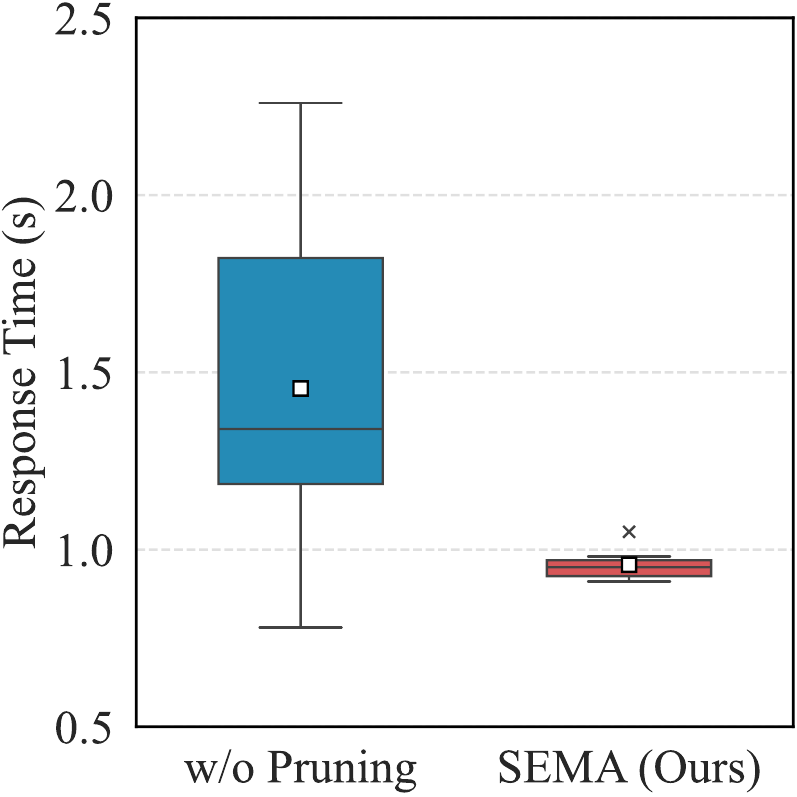}}\hfill
    \subfloat[8m]{\includegraphics[width=0.23\textwidth]{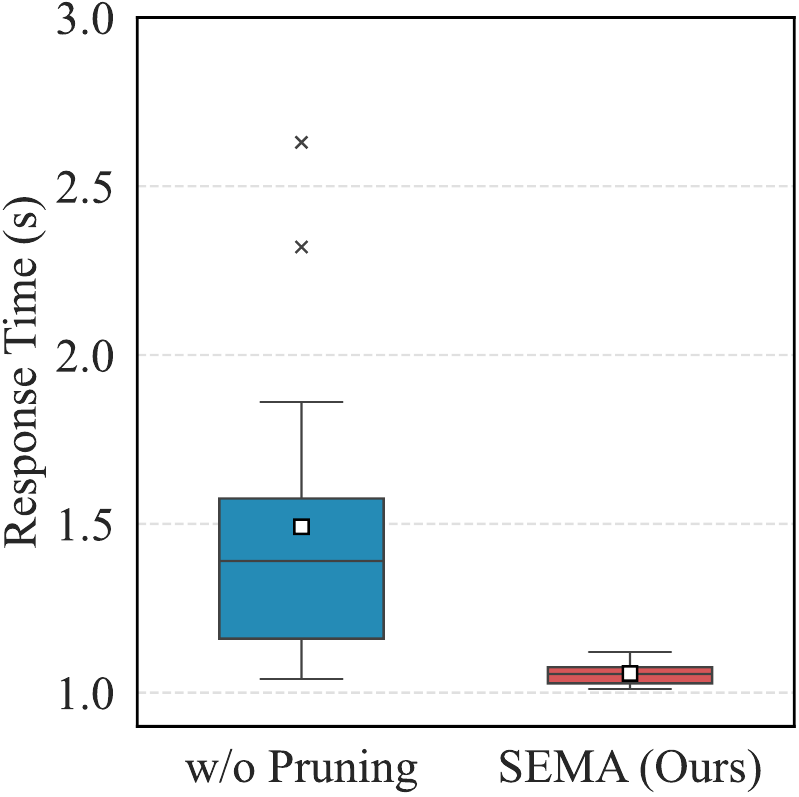}}\hfill
    \subfloat[Flat64-Lv.1]{\includegraphics[width=0.23\textwidth]{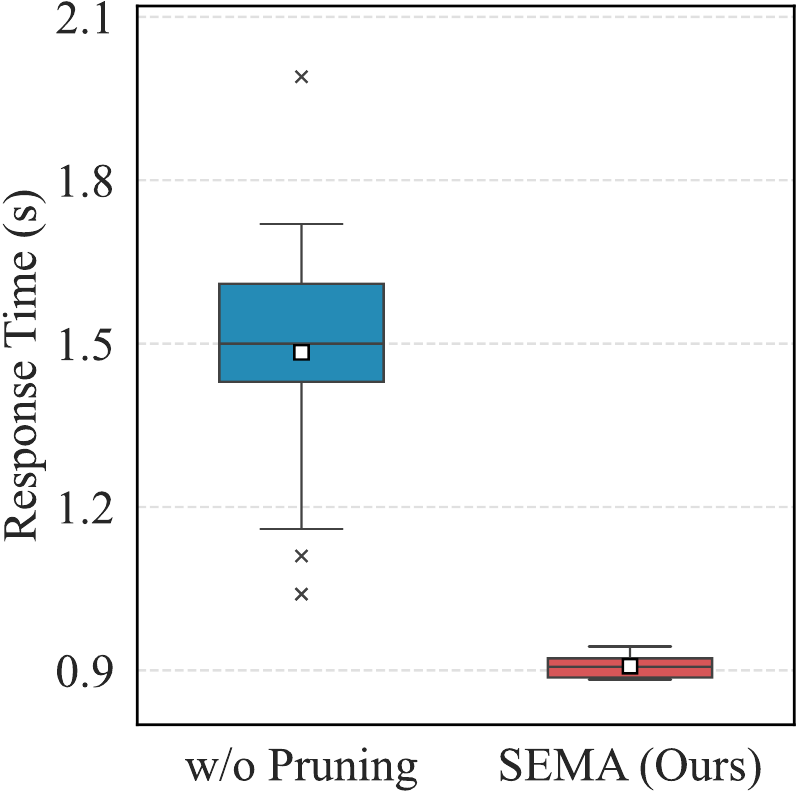}}\hfill
    \subfloat[Flat64-Lv.2]{\includegraphics[width=0.23\textwidth]{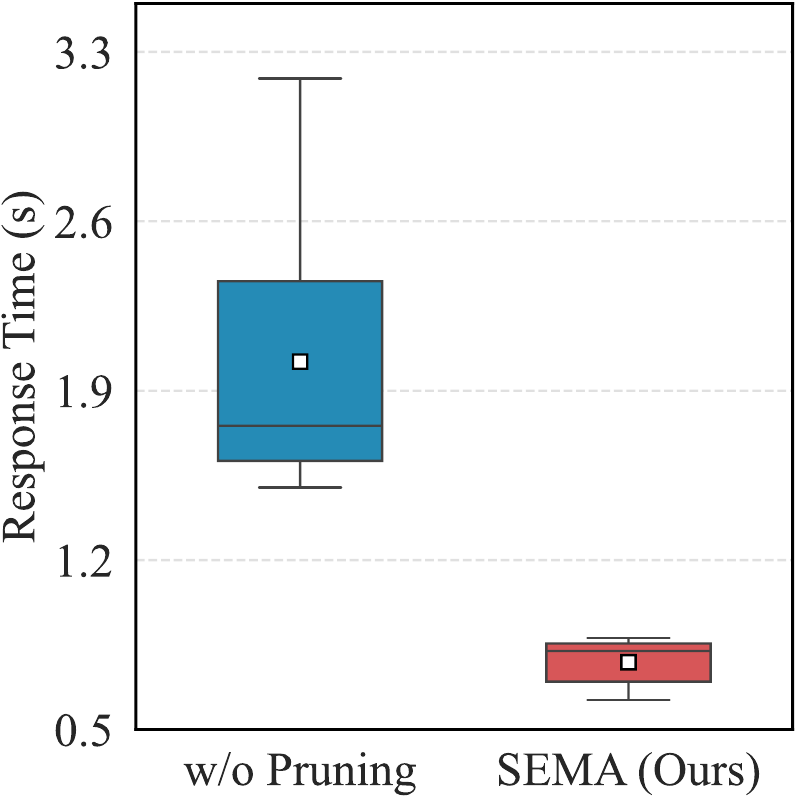}}
    \captionsetup{singlelinecheck=off, justification=raggedright}
    \caption{Ablation study of structural entropy driven pruning on response time across different maps and different levels. (a) 3m; (b) 8m; (c) Flat64(VeryEasy); (d) Flat64(Easy).}
    \label{fig4}
\end{figure*}

To rigorously verify the contributions of individual core components within the SEMA framework to overall system performance, we conducted a series of ablation studies. The experiments focused on three dimensions: the dynamic observation pruning mechanism, the Evaluation Agent, and the Policy Agent. Testing was performed across varying difficulty levels on SMAC micromanagement maps 3m and 8m, as well as Melee macro-strategy maps Flat64 and Simple64. The experimental results demonstrate that the synergy among these components is fundamental to achieving efficient and robust multi-agent decision-making.

Table \ref{tab3} summarizes the specific performance of each variant in terms of win rate and step response time. Particularly in the 3m and 8m maps, which demand extreme micromanagement, the average response time surged to over 1.0s. In the Simple64-Lv.2 macro map, latency even reached 2.1s. This high latency directly caused a precipitous drop in win rates; for instance, the win rate in the 3m map fell from 88\% to 50\%.

The decisive role of dynamic pruning on inference efficiency is clearly demonstrated by the data. Experimental results indicate that the dynamic pruning mechanism is critical for ensuring system real-time responsiveness. As illustrated in Figure 4, after removing the pruning module in the variant without Dynamic Pruning, system response time increased significantly across all test maps and exhibited severe fluctuations.  This validates our perspective that due to the inherent inference lag of LLMs when processing raw high-dimensional data, instructions cannot be delivered within effective observation windows without efficient information compression, leading to tactical failure.

Multi-agent collaboration serves as a guarantee for decision quality. Beyond the dimension of efficiency, the Evaluation Agent and Policy Agent are indispensable for maintaining high-level decision logic. Data in Table \ref{tab3} shows that after removing the Policy Agent in the variant without Policy Agent, the system win rate dropped from 70\% to 52\% in the 8m map, and from 100\% to 72\% in the Simple64-Lv.2 task. This suggests that the absence of post-match reflection and experience iteration causes the framework to lose its self-optimization capability, making it difficult to handle complex adversarial environments. Meanwhile, while removing the Evaluation Agent in the variant without Evaluation Agent had a minor impact on inference latency, which maintained a range between 0.6s and 0.7s, the lack of real-time experience retrieval and command correction led to noticeable stochastic errors and instability when facing high-difficulty AI at Lv.2, resulting in varying degrees of decline in win rates.

\subsection{Hyperparameter Sensitivity Analysis}

\begin{table*}[t!]
\centering
\caption{Analysis of information retention rates across different combat phases and hyperparameter configurations.}
\label{tab4}
\footnotesize
\begin{tabularx}{\textwidth}
    {
        >{\hsize=1.0\hsize}C 
        >{\hsize=1.0\hsize}C 
        *{6}{>{\hsize=1.0\hsize}C} 
    }
    \toprule
    Phase & ($\alpha, \beta$) & Resource & Building & Unit & Research & Planning & Total \\ 
    \midrule
    \multirow{2}{*}{Early} & (0.95, 0.05) & 36.6 & 75.9 & 63.6 & 76.4 & 0 & 19.1 \\
                           & (0, 1) & 53.8 & 59.5 & 55.8 & 81.2 & 0 & 18.3 \\
    \midrule
    \multirow{2}{*}{Mid}   & (0.95, 0.05) & 84.6 & 65.6 & 77.8 & 100 & 100 & 24.0 \\
                           & (0, 1) & 29.0 & 90.7 & 75.5 & 99.9 & 99.7 & 22.2 \\
    \midrule
    \multirow{2}{*}{Late}  & (0.95, 0.05) & 62.1 & 71.4 & 73.0 & 100 & 0 & 19.2 \\
                           & (0, 1) & 68.5 & 84.4 & 69.3 & 70.4 & 0 & 19.6 \\
    \bottomrule
\end{tabularx}
\end{table*}

\begin{figure*}[t!]
\centering
    \subfloat[\hspace{1em}the 30th Frame]{\includegraphics[width=0.26\textwidth]{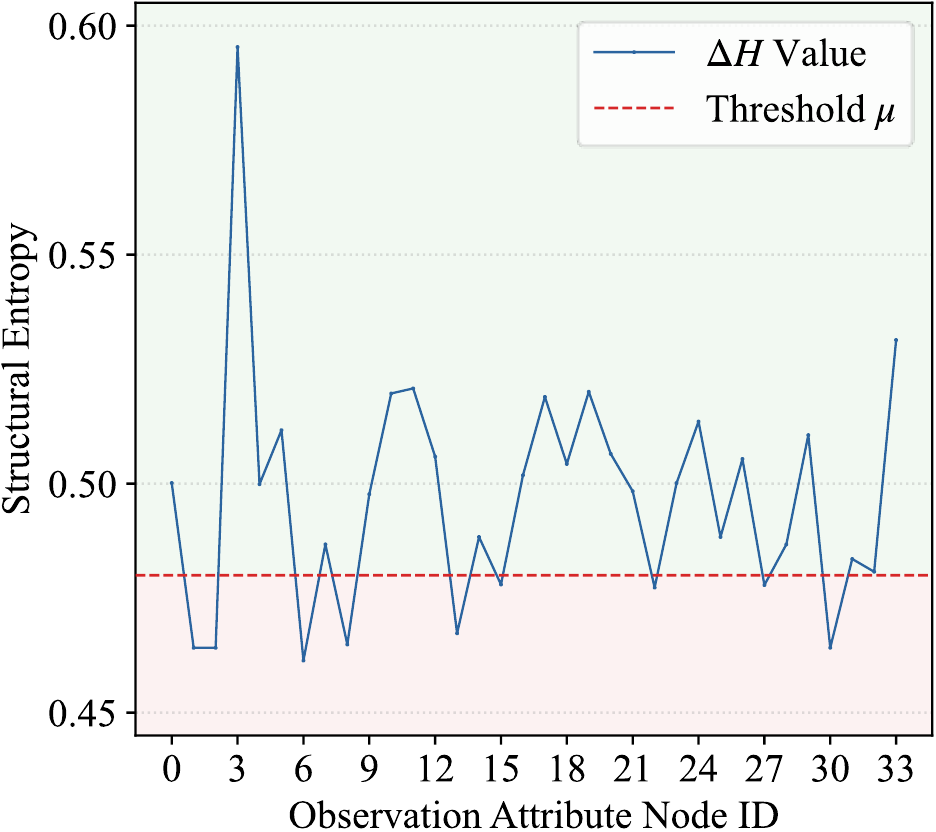}\label{fig5:1}}
    \hspace{0.25cm} 
    \subfloat[\hspace{1em}the 200th Frame]{\includegraphics[width=0.26\textwidth]{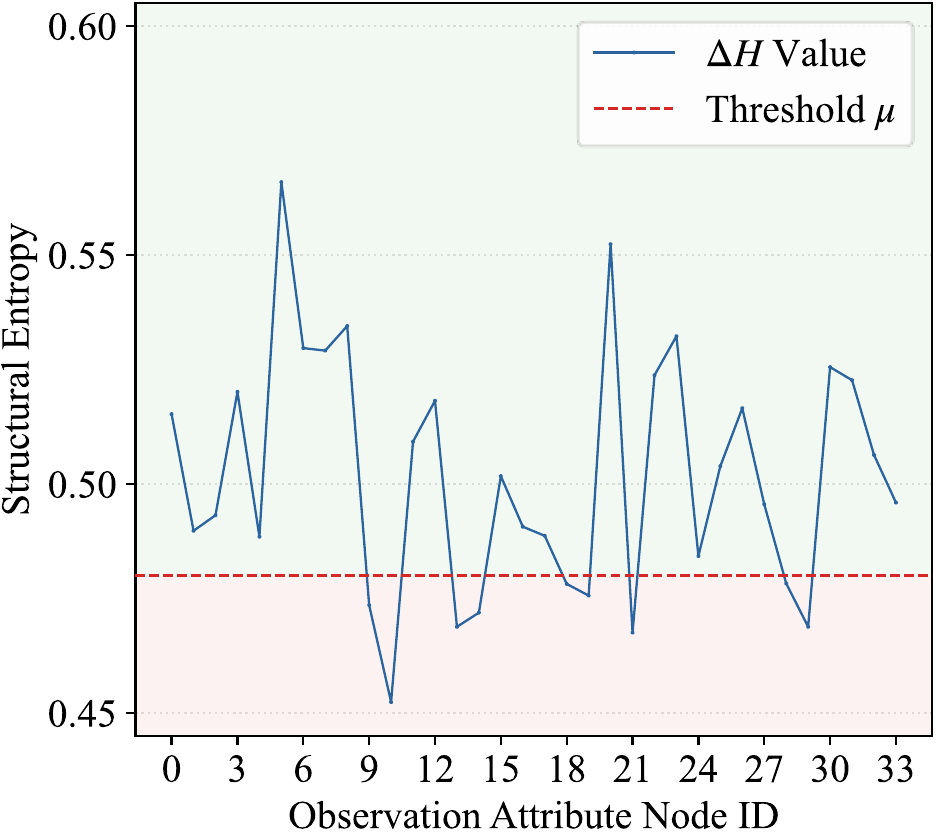}\label{fig5:2}}
    \hspace{0.25cm} 
    \subfloat[Impact of $N$ on Win Rate and Token Consumption]{\includegraphics[width=0.418\textwidth]{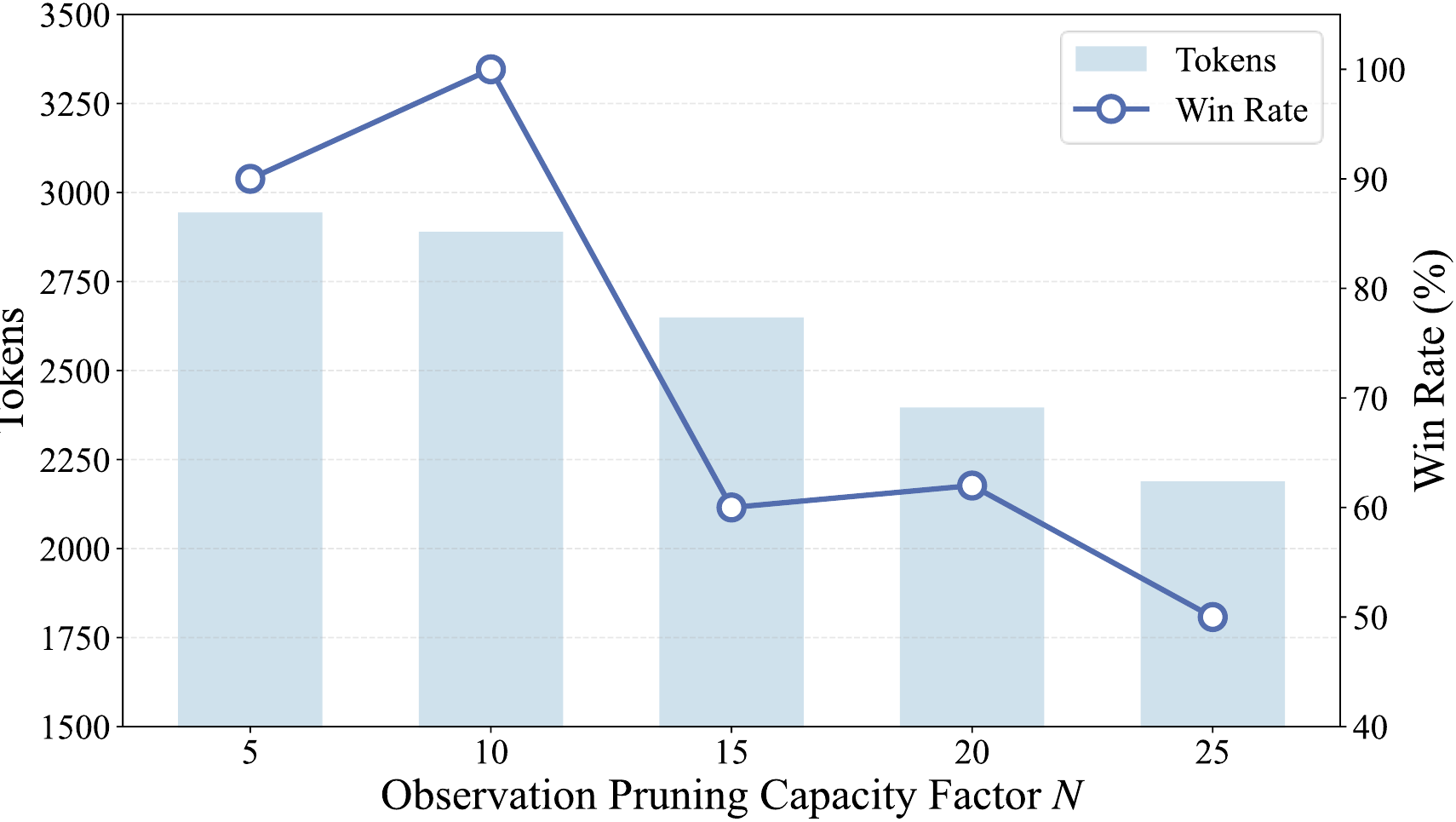}\label{fig5:3}}
    \captionsetup{singlelinecheck=off, justification=raggedright}
    \caption{Hyperparameter sensitivity analysis and semantic filtering visualization on the Simple64-Lv2 map. (a) and (b) (at frame 30 and frame 200, respectively) visualize the structural entropy-driven dynamic pruning process of observation information, where red regions indicate observation attribute nodes with $\Delta H$ below the threshold $\mu$ that require pruning, while green regions represent retained attribute nodes; (c) shows the relationship between average win rate and the number of tokens under different capacity factors $N$.}
    \label{fig5}
\end{figure*}

In this subsection, we analyze the impact of the dynamic observation pruning mechanism based on structural entropy and its key hyperparameters on framework performance. Specifically, the LLM first performs static filtering on the attribute information of the 105 observation dimensions. Then we conduct structural modeling and dynamic pruning based on the remaining 33 attributes. Figure \ref{fig5:1} and Figuire \ref{fig5:2} represent the 30th and 200th frames of a single episode on the Simple64-Lv2 map, respectively, illustrating the distribution of pruning structural entropy $\Delta H$ for attribute nodes across different combat stages. A threshold $\mu = 0.48$ is established as the criterion for semantic filtering; nodes with entropy values exceeding this threshold are defined as critical semantic nodes and retained, while low-entropy nodes are identified as redundant information and subsequently pruned. Furthermore, the entropy values of various nodes exhibit dynamic fluctuations as the battlefield situation evolves. This demonstrates that SEMA can identify shifts in information priority in real time, ensuring that the large language model consistently focuses on the core semantics most relevant to current decision-making.

The selection of the pruning capacity factor $N$ directly dictates the equilibrium between inference efficiency and decision quality. As shown in Figure \ref{fig5:3}, an increase in the value of $N$, representing enhanced pruning intensity, results in fewer retained information nodes. Consequently, the number of tokens consumed per inference decreases, accompanied by a gradual decline in win rates. Experimental results indicate that the framework achieves an optimal balance between information compression and semantic retention when $N=10$. Insufficient values of $N$ lead to the loss of critical observational data due to excessive pruning, whereas values exceeding 15 introduce redundant information that interferes with the strategic reasoning logic of the model, resulting in diminished win rates. This corroborates the necessity of moderate pruning to maintain the accuracy of high-level strategic reasoning.

Moreover, we investigate the influence of the weighting coefficients $\alpha$ and $\beta$ within the structural entropy formula $\delta_H = \alpha \cdot \Delta H_G + \beta \cdot \Delta H_T$ on the retention rates of various information types. As detailed in Table \ref{tab4}, a comparison between the configurations of $(\alpha, \beta)$ at $(0.95, 0.05)$ and $(1, 0)$ reveals that when the global weight $\alpha=0.95$ is dominant, the system prioritizes long-cycle categories such as Research and Planning, which reflect macro-level strategic trends. In contrast, when global topological entropy is entirely neglected in favor of local transformation weights i.e., $\alpha=0, \beta=1$, the focus shifts toward fundamental data with high-frequency numerical fluctuations, such as Resources and Buildings. Experimental evidence suggests that a tuning approach prioritizing global topological entropy supplemented by minor local weights enables SEMA to maintain a macro-level grasp of strategic evolution while preserving essential local dynamic perceptions, thereby sustaining a robust competitive advantage at minimal cost. Notably, the retention rate of Planning nodes exhibits a distinct phased characteristic, being fully retained only during the mid-game to support complex strategic reasoning. During the early stage, before strategies have solidified, or the late stage, when they have fully developed, these nodes are identified as low-entropy redundancies and pruned due to their contribution falling below the established threshold.

Finally, we conducted a evaluation of the decision-making overhead of different methods across various experimental maps. As shown in Table \ref{tab5}, while significantly improving decision quality, SEMA achieved extremely low input costs for the large model. Compared to baseline methods, SEMA reduced the input length required for each average decision step by an order of magnitude. In the most complex 25m map, HIMA's average token consumption per step was as high as 33.1k, while SEMA required only 2.2k; even in macro Melee maps, SEMA still stably controlled token consumption at around 2.9k, far lower than schemes such as Single-LLM and TextStarCraft. This fully proves that SEMA, by virtue of its structural entropy-driven dynamic pruning mechanism, significantly compresses observation redundancy under the premise of ensuring the integrity of core battlefield semantics, thereby achieving superior inference efficiency and extremely high computational economy in real-time combat.

\begin{table*}[t!]
\centering
\caption{Comparative analysis of token consumption per step across different scenarios.}
\label{tab5}
\footnotesize
\begin{tabularx}{\textwidth}
    {
        >{\hsize=1.4\hsize}C 
        *{3}{>{\hsize=0.6\hsize}C} 
        *{6}{>{\hsize=0.65\hsize}C} 
    }
    \toprule
    \multirow{2}{*}{Method} & \multirow{2}{*}{3m} & \multirow{2}{*}{8m} & \multirow{2}{*}{25m} & \multicolumn{2}{c}{Flat32} & \multicolumn{2}{c}{Flat48} & \multicolumn{2}{c}{Simple64} \\ 
    \cmidrule(lr){5-10}
    & & & & Lv.1 & Lv.2 & Lv.1 & Lv.2 & Lv.1 & Lv.2\\
    \midrule
    Single-LLM    & 7.0k & 8.3k & 12.1k &  14.0k & 16.4k & 12.1k& 12.2k & 12.5k & 13.3k \\
    TextStarCraft & 2.9k & 3.4k & 5.1k  & 15.9k & 15.4k & 15.2k & 15.0k & 15.2k & 15.6k \\
    HIMA          & 12.1k & 13.9k & 33.1k & 11.1k & 11.4k & 12.1k & 12.8k & 10.4k & 11.8k \\
    \textbf{SEMA (Ours)} & \textbf{0.9k} & \textbf{1.5k} & \textbf{2.2k} & \textbf{2.9k} & \textbf{2.9k}  & \textbf{2.9k} & \textbf{2.8k} & \textbf{2.9k} & \textbf{2.9k} \\
    \bottomrule
\end{tabularx}
\end{table*}

\subsection{Case Study}
\begin{figure*}[t!]
    \centering
    \subfloat[Snapshot of Episode 2 \label{fig6:1}]{
        \includegraphics[width=0.333\textwidth]{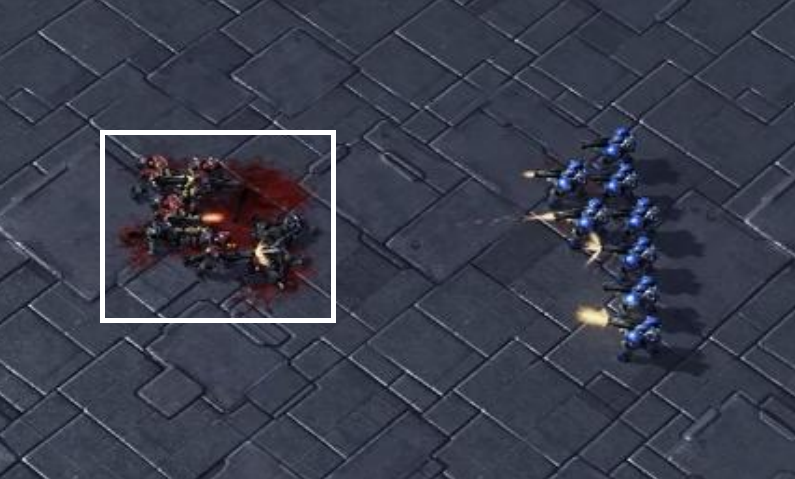}
    }
    \subfloat[Snapshot of Episode 3 \label{fig6:2}]{
        \includegraphics[width=0.31\textwidth]{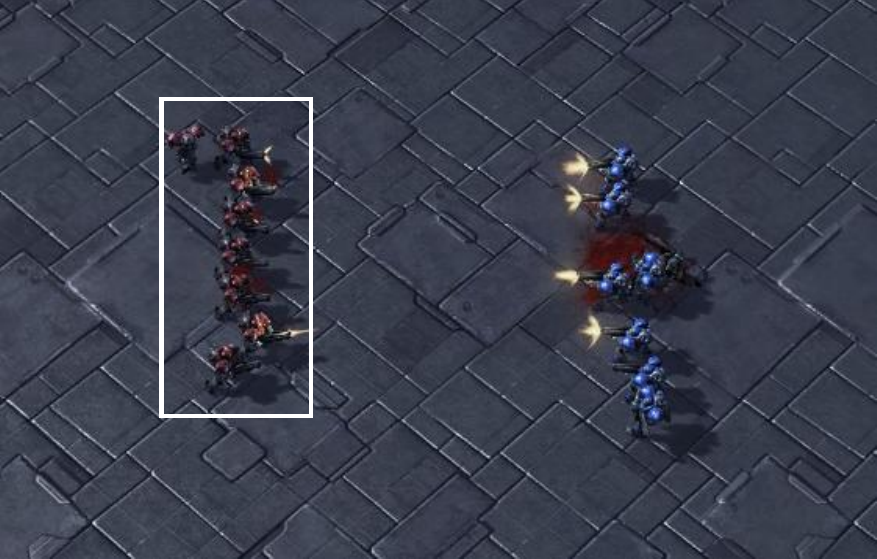}
    }
    \subfloat[Snapshot of Episode 4 \label{fig6:3}]{
        \includegraphics[width=0.31\textwidth]{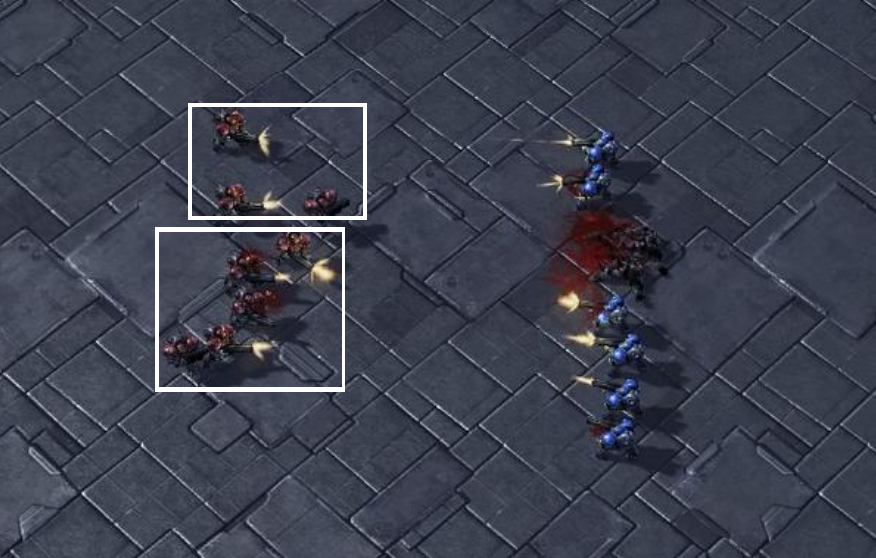}
    }
    \captionsetup{singlelinecheck=off, justification=raggedright}
    \caption{Snapshots of tactical evolution on the 8m map ($\mu = 0.48$). (a) illustrates the disordered state in Episode 2, while (b) demonstrates the collaborative formation and focus fire strategy in Episode 3.} 
    \label{fig:snapshots_evolution}
\end{figure*}
To verify the self-evolving capability of the SEMA framework in dynamic adversarial environments, this study selects typical tactical performances from three consecutive game episodes on the 8m map for comparative analysis.

In the Episode 2 shown in Figure \ref{fig6:1}, the policy agent is in an early exploration stage. Units rely on local observations, leading to disordered spatial distribution and dispersed firepower. This lack of coordination results in significant numerical inferiority and unit losses, highlighting the limitations of the initial policy in complex collaborative tasks. To address these deficiencies, the Policy Agent proposes refined strategies aimed at enhancing focus-fire efficiency and maximizing effective firepower coverage.

In contrast, Episode 3 demonstrates strategic evolution via failure-based correction. Leveraging spatial data from SEMA, the agent reconstructs structured tactical mappings. As illustrated in Figure \ref{fig6:2}, units transition into a regular linear formation upon engagement to maximize firepower. By executing sub-group focus fire consisting of three to four units, the agent successfully reverses the loss ratio through superior tactical coordination.

However, the linear formation still suffers from excessive clustering and a lack of effective kiting. To address this, the Policy Agent formulates a strategy to prevent clumping and establish efficient grouping. In Episode 4, the system exhibits advanced tactical division. The group implements dynamic resource allocation based on enemy positioning. As shown in Figure \ref{fig6:3}, our forces decouple into two modules: a 3-Marine flanking group suppressing northern targets, and a 5-unit primary force concentrating fire on southern targets. This "divide-and-conquer" approach significantly enhances kill efficiency while maintaining formation stability.

This case intuitively proves the self-evolution capability of SEMA. The Policy Agent is able to analyze strengths and weaknesses in every match, ultimately evolving highly competitive tactical behaviors in complex RTS games.

\vspace{10pt}
\noindent
\begin{minipage}[t]{0.48\textwidth}
    \begin{promptbox}{Policy Evolution for Episode 2}
        \small
        Poor focus fire reflects firepower dispersion. By 00:04, friendly units sustained severe damage, indicating failed prioritization. By 00:08, only 5 units remained. To improve, a structured formation is required to concentrate fire output and maximize the global firepower coverage area effectively and consistently.
    \end{promptbox}
\end{minipage}
\hfill
\begin{minipage}[t]{0.48\textwidth}
    \begin{promptbox}{Policy Evolution for Episode 3}
        \small
        Despite equal initial counts, the friendly side suffered two rapid losses. Our units exhibit excessive clustering and a collapsed formation. To improve, structured grouping is required to facilitate tactical division, allowing different formations to engage distinct enemy targets while maintaining superior focus fire efficiency.
    \end{promptbox}
\end{minipage}

\section{Conclusion}
To address the core challenges of state-space explosion and real-time response latency faced by LLMs in RTS scenarios, We propose SEMA, a self-evolving multi-agent collaborative framework.

First, at the perceptual level, this study bypasses traditional raw data input by implementing a dynamic observation pruning mechanism driven by structural entropy. Through topological modeling and semantic compression, this mechanism reduces the LLM input load by exceeding 70\% while retaining core strategic semantics, thereby resolving decision latency caused by high-dimensional redundancy. Second, at the decision evolution level, a closed-loop system of decision, evaluation, and analysis agents was constructed. By integrating trajectory memory retrieval and global empirical reflection, the system achieves coverage from instantaneous action calibration to long-term strategic evolution, suppressing LLM stochastic bias and enabling self-correction in complex adversarial environments.

Experimental results demonstrate that SEMA exhibits superior competitive performance across a variety of complex maps in StarCraft II. Compared to current mainstream LLM decision-making schemes, SEMA achieves a significant advantage in win rates, far surpassing other baseline methods. The average decision response time is controlled within 1.0 second. This framework provides a novel technical path for autonomous agent decision-making in constrained urban environments or real-time adversarial scenarios. Future research will further explore more robust cross-domain strategy transfer algorithms to address increasingly extreme non-cooperative game environments.




\begin{thebibliography}{1}




\bibitem{1} Silver D, Huang A, Maddison C J, et al. Mastering the game of Go with deep neural networks and tree search[J]. nature, 2016, 529(7587): 484-489.

\bibitem{2} Park J S, O'Brien J, Cai C J, et al. Generative agents: Interactive simulacra of human behavior[C]//Proceedings of the 36th annual acm symposium on user interface software and technology. 2023: 1-22.

\bibitem{3} Vinyals O, Ewalds T, Bartunov S, et al. Starcraft ii: A new challenge for reinforcement learning[J]. arXiv preprint arXiv:1708.04782, 2017.

\bibitem{4} Vinyals O, Babuschkin I, Czarnecki W M, et al. Grandmaster level in StarCraft II using multi-agent reinforcement learning. Nature, 2019, 575: 350--354.

\bibitem{5} Wang G, Xie Y, Jiang Y, et al. Voyager: An open-ended embodied agent with large language models[J]. arXiv preprint arXiv:2305.16291, 2023.

\bibitem{6} Yao S, Yu D, Zhao J, et al. Tree of thoughts: Deliberate problem solving with large language models. In: Proceedings of NeurIPS, New Orleans, 2023. 11809--11822.

\bibitem{7} Brown T, Mann B, Ryder N, et al. Language models are few-shot learners. In: Proceedings of NeurIPS, 2020. 1877--1901.

\bibitem{8} Ontanón S, Synnaeve G, Uriarte A, et al. A survey of real-time strategy game AI research and competition in StarCraft. IEEE Transactions on Computational Intelligence and AI in Games, 2013, 5(4): 293--311.

\bibitem{9} Silver D, Hubert T, Schrittwieser J, et al. A general reinforcement learning algorithm that masters chess, shogi, and Go through self-play. Science, 2018, 362(6419): 1140--1144. 

\bibitem{10} Hao S, Gu Y, Ma H, et al. Reasoning with language model is planning with world model[C]//Proceedings of the 2023 Conference on Empirical Methods in Natural Language Processing. 2023: 8154-8173.

\bibitem{11} Ahn D, Kim S, Choi J. Society of Mind Meets Real-Time Strategy: A Hierarchical Multi-Agent Framework for Strategic Reasoning[J]. arXiv preprint arXiv:2508.06042, 2025.

\bibitem{12} Park G, Lee S, Park Y. Minimizing Response Latency in LLM-Based Agent Systems: A Comprehensive Survey[J]. IEEE Access, 2026.

\bibitem{13} Shao H, Hu Y, Wang L, et al. Lmdrive: Closed-loop end-to-end driving with large language models[C]//Proceedings of the IEEE/CVF conference on computer vision and pattern recognition. 2024: 15120-15130.

\bibitem{14} Arulkumaran K, Cully A, Togelius J. Alphastar: An evolutionary computation perspective[C]//Proceedings of the genetic and evolutionary computation conference companion. 2019: 314-315.

\bibitem{15} Wang X, Wei J, Schuurmans D, et al. Self-consistency improves chain of thought reasoning in language models[J]. arXiv preprint arXiv:2203.11171, 2022.

\bibitem{16} Ahn D, Kim S, Choi J. Society of Mind Meets Real-Time Strategy: A Hierarchical Multi-Agent Framework for Strategic Reasoning[J]. arXiv preprint arXiv:2508.06042, 2025.

\bibitem{17} Wooldridge M, Jennings N R. Intelligent agents: Theory and practice[J]. The knowledge engineering review, 1995, 10(2): 115-152.

\bibitem{18} Hong S, Zhuge M, Chen J, et al. MetaGPT: Meta programming for a multi-agent collaborative framework[C]//The twelfth international conference on learning representations. 2023.

\bibitem{19} Qian C, Liu W, Liu H, et al. Chatdev: Communicative agents for software development[C]//Proceedings of the 62nd Annual Meeting of the Association for Computational Linguistics (Volume 1: Long Papers). 2024: 15174-15186.

\bibitem{20} Li G, Hammoud H, Itani H, et al. Camel: Communicative agents for" mind" exploration of large language model society[J]. Advances in Neural Information Processing Systems, 2023, 36: 51991-52008.

\bibitem{21} Gong R, Huang Q, Ma X, et al. Mindagent: Emergent gaming interaction[C]//Findings of the Association for Computational Linguistics: NAACL 2024. 2024: 3154-3183.

\bibitem{22} Park J S, O'Brien J, Cai C J, et al. Generative agents: Interactive simulacra of human behavior[C]//Proceedings of the 36th annual acm symposium on user interface software and technology. 2023: 1-22.

\bibitem{23} Zeng Y, Li S, Li Z, et al. Communication-Efficient Collaborative Perception with Semantic and Statistical Compression[C]//Chinese Conference on Pattern Recognition and Computer Vision (PRCV). Singapore: Springer Nature Singapore, 2025: 270-283.

\bibitem{24} Zhang Y, Ma Z, Ma Y, et al. Webpilot: A versatile and autonomous multi-agent system for web task execution with strategic 

\bibitem{25} Zhang X D, Xiao F, Zheng Y S, et al. Asynchronous and aperiodic sampled-data control for general linear multiagent systems. Sci China Inf Sci, 2026, 69(2): 122204, https://doi.org/10.1007/s11432-024-4391-2

\bibitem{26} Wei J, Wang X, Schuurmans D, et al. Chain-of-thought prompting elicits reasoning in large language models[J]. Advances in neural information processing systems, 2022, 35: 24824-24837.

\bibitem{27} Yao S, Yu D, Zhao J, et al. Tree of thoughts: Deliberate problem solving with large language models[J]. Advances in neural information processing systems, 2023, 36: 11809-11822.

\bibitem{28} Besta M, Blach N, Kubicek A, et al. Graph of thoughts: Solving elaborate problems with large language models[C]//Proceedings of the AAAI conference on intelligence. 2024, 38(16): 17682-17690.

\bibitem{29} Kim M J, Pertsch K, Karamcheti S, et al. Openvla: An open-source vision-language-action model. arXiv preprint arXiv:2406.09246, 2024.

\bibitem{30} Lei M, Wang G, Zhao Y, et al. Clea: Closed-loop embodied agent for enhancing task execution in dynamic environments. In: Proceedings of the 2025 IEEE/RSJ International Conference on Intelligent Robots and Systems (IROS). IEEE, 2025. 21048-21054.

\bibitem{31} Wang Z, Cai S, Chen G, et al. Describe, explain, plan and select: interactive planning with llms enables open-world multi-task agents[J]. Advances in Neural Information Processing Systems, 2023, 36: 34153-34189.

\bibitem{32} Silver D, Huang A, Maddison C J, et al. Mastering the game of Go with deep neural networks and tree search[J]. nature, 2016, 529(7587): 484-489.

\bibitem{33} Valmeekam K, Marquez M, Olmo A, et al. Planbench: An extensible benchmark for evaluating large language models on planning and reasoning about change[J]. Advances in Neural Information Processing Systems, 2023, 36: 38975-38987.

\bibitem{34} Wang Y, Ji P, Yang C, et al. Mcts-judge: Test-time scaling in llm-as-a-judge for code correctness evaluation[J]. arXiv preprint arXiv:2502.12468, 2025.

\bibitem{35} Hao S, Gu Y, Ma H, et al. Reasoning with language model is planning with world model[C]//Proceedings of the 2023 Conference on Empirical Methods in Natural Language Processing. 2023: 8154-8173.

\bibitem{36} Saha S, Li X, Ghazvininejad M, et al. Learning to plan \& reason for evaluation with thinking-llm-as-a-judge[J]. arXiv preprint arXiv:2501.18099, 2025.

\bibitem{37} Xi Z, Chen W, Guo X, et al. The rise and potential of large language model based agents: A survey[J]. Science China Information Sciences, 2025, 68(2): 121101.

\bibitem{38} Wan G, Wu Y, Chen J, et al. Reasoning aware self-consistency: Leveraging reasoning paths for efficient llm sampling[C]//Proceedings of the 2025 Conference of the Nations of the Americas Chapter of the Association for Computational Linguistics: Human Language Technologies (Volume 1: Long Papers). 2025: 3613-3635.

\bibitem{39} Vinyals O, Babuschkin I, Czarnecki W M, et al. Grandmaster level in StarCraft II using multi-agent reinforcement learning[J]. nature, 2019, 575(7782): 350-354.

\bibitem{40} Wang G, Xie Y, Jiang Y, et al. Voyager: An open-ended embodied agent with large language models[J]. arXiv preprint arXiv:2305.16291, 2023.

\bibitem{41} Lin H, Shi Y, Geng T, et al. Agent-Omni: Test-Time Multimodal Reasoning via Model Coordination for Understanding Anything[J]. arXiv preprint arXiv:2511.02834, 2025.

\bibitem{42} Sima C, Renz K, Chitta K, et al. Drivelm: Driving with graph visual question answering[C]//European conference on computer vision. 2024.

\bibitem{43} Zhao Y, Li Z, Jin Z, et al. PGPL: enhancing spatial awareness abilities of multimodal large language models based on precise geometric position learning[J]. Science China Information Sciences, 2026, 69(2): 1-17

\bibitem{44} Wang H, Wang J, Leong C T, et al. Steca: Step-level trajectory calibration for llm agent learning[C]//Findings of the Association for Computational Linguistics: ACL 2025. 2025: 11597-11614

\bibitem{45} Xu J, Nan G, Guan S, et al. Refining positive and toxic samples for dual safety self-alignment of llms with minimal human interventions[J]. IEEE Transactions on Information Forensics and Security, 2026.

\bibitem{46} Mozikov M, Severin N, Bodishtianu V, et al. Eai: Emotional decision-making of llms in strategic games and ethical dilemmas[J]. Advances in Neural Information Processing Systems, 2024, 37: 53969-54002.

\bibitem{47} Zhang S, Liang J, Zhou Z, et al. ORION: Option-Regularized Deep Reinforcement Learning for Cooperative Multi-Agent Online Navigation[J]. arXiv preprint arXiv:2601.01155, 2026.

\bibitem{48} Xue D, Zhou X J, Wang M, et al. Formation control and path planning of multi-robot systems via large language models. Sci China Inf Sci, 2025, 68(5): 150205, https://doi.org/10.1007/s11432-024-4290-4

\bibitem{49} Zhang Y, Jiang S, Li R, et al. DeepPlanning: Benchmarking Long-Horizon Agentic Planning with Verifiable Constraints[J]. arXiv preprint arXiv:2601.18137, 2026.

\bibitem{50} Ma W, Mi Q, Zeng Y, et al. Large language models play starcraft ii: Benchmarks and a chain of summarization approach[J]. Advances in Neural Information Processing Systems, 2024, 37: 133386-133442.

\bibitem{51} Li A, Pan Y. Structural information and dynamical complexity of networks[J]. IEEE Transactions on Information Theory, 2016, 62(6): 3290-3339.

\bibitem{52} Oliehoek F A, Amato C. A concise introduction to decentralized POMDPs[M]. Cham, Switzerland: Springer International Publishing, 2016.

\bibitem{53} Yang A, Li A, Yang B, et al. Qwen3 technical report[J]. arXiv preprint arXiv:2505.09388, 2025.

\end{thebibliography}

\end{document}